\documentclass{article}
\usepackage[letterpaper]{geometry}
\usepackage{amssymb,amsfonts}
\usepackage{graphicx}
\usepackage{mathtools}
\usepackage{xfrac}
\usepackage{hyperref}
\usepackage{cite}
\graphicspath{{{./}}}

\DeclareFontFamily{OT1}{pzc}{}
\DeclareFontShape{OT1}{pzc}{m}{it}{<-> s * [1.10] pzcmi7t}{}
\DeclareMathAlphabet{\mathpzc}{OT1}{pzc}{m}{it}

\newcommand{\aref}[1]{App.\,\ref{#1}}
\newcommand{\fref}[1]{Fig.\,\ref{#1}}
\newcommand{\tref}[1]{Table \ref{#1}}
\newcommand{\eref}[1]{Eq.\,(\ref{#1})}

\newcommand{\sref}[1]{Sec.\!~\ref{#1}}

\newcommand{\cref}[1]{Ref.\,\cite{#1}}
\newcommand{\crefs}[1]{Refs.\,\cite{#1}}

\newcommand{\vs}{{\it vs.}\! }
\newcommand{\ie}{{\it i.e.}\! }
\newcommand{\eg}{{\it e.g.}\! }

\newcommand{\etal}{{\it et al.}\! }

\newcommand{\aposteriori}{{\it a posteriori} }

\newcommand{\pdf}{PDF}
\newcommand{\prob}{p} 

\newcommand{\Zc}{\mathcal{Z}}
\newcommand{\Yc}{\mathcal{Y}}

\newcommand{\Cbb}{\mathbb{C}}

\newcommand{\xib}{{\boldsymbol{\xi}}}

\newcommand{\epsilonb}{\boldsymbol{\epsilon}}
\newcommand{\sigmab}{\boldsymbol{\sigma}}

\renewcommand{\sb}{\mathbf{s}}

\newcommand{\xb}{\mathbf{x}}

\newcommand{\Ib}{\mathbf{I}}

\newcommand{\tr}{\operatorname{tr}}

\newcommand{\dev}{\operatorname{dev}}

\newcommand{\PHYSICAL}{{\boldsymbol{\theta}}}
\newcommand{\KLE}{\boldsymbol{\vartheta}}
\newcommand{\EMBEDDED}{\boldsymbol{\lambda}}
\newcommand{\IIDGERMS}{\boldsymbol{\xi}}
\newcommand{\DELTA}{\boldsymbol{\Delta}}
\newcommand{\HYPER}{\boldsymbol{\Gamma}}
\newcommand{\hyper}{\gamma}
\newcommand{\bhyper}{\boldsymbol{\gamma}}
\newcommand{\FEATURES}{\mathcal{F}}
\newcommand{\DATA}{\mathsf{D}}
\newcommand{\noise}{\eta}
\newcommand{\NOISE}{\boldsymbol{\eta}}
\newcommand{\feature}{f}
\newcommand{\expectation}{\mathbb{E}}
\newcommand{\bd}{\boldsymbol{d}}
\newcommand{\meanporosity}{\bar{\varphi}}
\newcommand{\porosity}{{\varphi}}

\title{\bf Modeling strength and failure variability due to porosity in additively manufactured metals}
\author{
M. Khalil$^\ddagger$,
G. H. Teichert$^\sharp$,
C. Alleman$^\ddagger$,
N.M. Heckman$^\dagger$,\\
R.E. Jones$^\ddagger$\thanks{corresponding: rjones@sandia.gov},\,
K. Garikipati$^\sharp$,
B.L. Boyce$^\dagger$\\[0.1in]
$^\ddagger${\it Sandia National Laboratories, Livermore, CA 94551}\\
$^\sharp${\it University of Michigan, Ann Arbor, MI 48109}\\
$^\dagger${\it Sandia National Laboratories, Albuquerque, NM 87185}\\
}
\date{}

\begin{document}
\maketitle

\begin{abstract}
To model and quantify the variability in plasticity and failure of additively manufactured metals due to imperfections in their microstructure, we have developed uncertainty quantification methodology based on pseudo marginal likelihood and embedded variability techniques.
We account for both the porosity resolvable in computed tomography scans of the initial material and the sub-threshold distribution of voids through a physically motivated model.
Calibration of the model indicates that the sub-threshold population of defects dominates the yield and failure response.
The technique also allows us to quantify the distribution of material parameters connected to microstructural variability created by the manufacturing process, and, thereby, make assessments of material quality and process control.
\end{abstract}

\section{Introduction} \label{sec:introduction}

Many materials, such as natural and additively manufactured (AM) materials, have widely varying properties due to variations in microstructure.
To predict the response of these technologically relevant materials and provide robust safety estimates and designs, the microstructural effects on the response must be accounted for in tractable numerical models.
Furthermore, representative models allow for inference of the sources of variability and structure-property relationships.
In this work, we present an uncertainty quantification based modeling technique enabled by multiscale simulation focused on AM metals.

AM components have advantages over conventionally fabricated parts.
AM enables complex geometries, rapid prototyping, and economical single and small batch builds, and yet AM materials tend to have significant variability due to the current state of the manufacturing process.
Sources of mechanical variability in AM materials are multifold: porosity, residual stresses, sintered interfaces, grain structures, surface imperfections, and gross deviations in geometry, and yet are primarily exaggerated versions of what can be found in traditionally fabricated parts.

In this work we focus on the porosity and damage aspects of the mechanical response of these materials.
We use high-throughput experimental stress-strain data and three dimensional highly resolved computed tomography (CT) scans that reveal widely varying failure characteristics and pervasive porosity in 17-4PH stainless steel dogbone-shaped test samples.
We model this AM material via realizations of the sample geometry on a finite element mesh with explicit representation of the observable voids together with a background density of unresolved voids via the selected constitutive damage model.
The effects of the sub-grid porosity are treated with the established approach of applying damage evolution models for ductile failure \cite{cocks1980intergranular,horstemeyer1999void,nahshon2008modification} to model void nucleation and growth.
Where the separation of scales assumed in this approach is violated by the larger defects, the voids are explicitly resolved by the topology of the finite element mesh, and their evolution is governed by the bulk transport of material dictated by the balance laws and constitutive models.
This hybrid approach has been found to be effective, for example, in evaluating the coupled effects of fine-scale voids and bulk defects \cite{karlson2019sandia} and in modeling the effects of particle cracking and ductile damage evolution in dual-phase materials \cite{paquet2011microstructural}.
It also provides a straightforward and expedient method for incorporating the experimental evidence available in this study.
There are many aspects of damage evolution, such as the detailed phenomenology of advancing cracks, that are not addressed by this type of approach, but the simulated failure behavior is qualitatively similar to that of real ductile materials, and good quantitative agreement will be demonstrated with respect to modeling variability of macroscopic performance metrics such as strain-to-failure.

The explicit characterization of porosity in AM materials for subsequent uncertainty propagation necessitates a stochastic description.
Quiblier \cite{Quiblier:1984} proposed a method that transforms white noise process realizations into discrete process counterparts, matching the target marginal probability density function (PDF) and autocorrelation function that characterize the porous media.
Adler \etal \cite{Adler:1990} subsequently extended that method to handle periodic boundary conditions.
An alternative strategy with ties to simulated annealing was proposed by Yeong and Torquato \cite{Yeong:1998} in which a random initial binary process realization is iteratively morphed into one that matches the target two-point correlation function.
Those approaches, albeit being able to handle various types of autocorrelation functions as well as anisotropy, are purely computational.
Recently, Ilango \etal \cite{Ilango:2016} developed an approach that constructs Karhunen-Lo\`eve expansion (KLE, see \eg \cref{le2010spectral}) models for 2-dimensional porosity processes.
KLE, also known as proper orthogonal decomposition or principal component analysis, is an efficient, reduced-order representation of a stochastic process that relies on the spectral decomposition of the corresponding correlation function.
This optimality in KLE representation (\ie capturing the variance of the process with the least number of modes) is essential in facilitating uncertainty propagation for subsequent Bayesian calibration by minimizing the number of random coefficients in representing porosity process.
We will extend this methodology to represent the 3-dimensional AM porosity processes as viewed from the CT data.

In this investigation, we are ultimately interested in inferring the physical parameters of the constitutive damage model so as to make robust predictions of performance and conjectures about the manufacturing process.
In order to handle various forms of error/uncertainties (\eg measurement noise, porosity variability), we adopt the Bayesian approach for statistical inference of the unknown physical parameters pertaining to the constitutive damage model.
In this context, the set of random coefficients that are used to model the high-dimensional (in terms of the number of random coefficients of the KLE) porosity process act as ``nuisance'' parameters which must be incorporated in the Bayesian calibration exercise.
The classical approach of jointly inferring those unknown coefficients is not feasible in this case due to the extreme dimensionality ($10^4$--$10^5$) of the nuisance parameter vector.
Instead, we will treat the observable porosity field as an uncontrollable or irreducible source of uncertainty in the physical system.
In uncertainty quantification (UQ) terms, such a source of uncertainty contributes to {\it aleatory} uncertainty \cite{Bedford:2001}, arising from the inherent variability of a phenomenon (in this case the random porosity distribution) which cannot be further reduced from the available data.
In contrast, the unknown parameter vector of interest contributes to epistemic uncertainty \cite{Bedford:2001} arising from incomplete knowledge of the phenomenon and can be reduced with the available data.
In this context, both types of uncertainty have probabilistic characterization while requiring separate treatment in a Bayesian setting (see \cref{Winkler:1996} for a relevant discussion).
We will formulate an approximation to the marginal likelihood of the physical parameters which marginalizes (integrates out) the random porosity process coefficients using a Monte Carlo (MC) method.
The resulting approximation falls under a general class of Pseudo-marginal Metropolis–Hastings methods \cite{Andrieu:2009}, with our specific approach being a variation of the Monte Carlo within Metropolis (MCWM) algorithm \cite{ONeill:2000} since it also relies on Metropolis-Hastings Markov chain Monte Carlo (MCMC) sampling \cite{gamerman2006markov,berg2008markov} of the physical parameters according to the resulting marginal posterior PDF.

Applications of Bayesian inference for model calibration often involve computational models that are assumed to be sufficiently consistent with reality.
In this context, the chosen constitutive damage model (described in \sref{sec:simulation}) can sufficiently capture the quantities of interest (QoIs), such as ultimate strength and failure strain, relating to plasticity and failure of AM metals on average across specimens or for specific specimens alone.
We will show that the physical model with nominal parameters however, does not fully capture the observed variability in the QoIs for the ensemble of specimens even with explicit porosity from a calibrated process.
In order to enhance the predictive accuracy and precision of calibrated constitutive models, we will capture the observed variability using model-form error to be calibrated in a Bayesian setting.
There are two popular approaches with Bayesian interpretation, the first of which employs an additive Gaussian process model discrepancy term for each observed QoI, as first proposed by Kennedy and O'Hagan \cite{kennedy2001bayesian}.
Despite its successful application in many contexts, modelers may prefer to adopt alternative approaches since (a) there is no guarantee that the predicted response will satisfy the underlying governing equations with all probability (\ie for all possible realizations) and (b) the additive error model is only able to provide predictive variability on the corresponding QoI that it was calibrated with (\ie it is unable to extrapolate variability to other QoIs).
For those reasons, another approach relying on the embedding of stochastic variables into the governing equations has recently gained momentum with the aim of capturing the observed variability with calibrated distributions of the stochastic variables \cite{Soize:2000,Strong:2014,sargsyan2015statistical,He:2016,Pernot:2017,Zio:2018,Morrison:2018,Hakim:2018}.
We will employ the framework first proposed by Sargsyan \etal \cite{sargsyan2015statistical} which embeds the variability in the model parameters with a parametric stochastic representation which is subsequently calibrated using the available observations.
We have used this approach in a previous study \cite{rizzi2019bayesian}, and, in this investigation, we build and expand upon that work in the richer context of failure and observed porosity.
In this work, we rely on the explicit porosity model as well as observational error model to capture, in part, the total observed variability.

In addition to our previous investigation \cite{rizzi2019bayesian}, the proposed framework relates to other works with probabilistic modeling applied to solid mechanics.
For example, Emery \etal \cite{emery2015predicting} constructed surrogate models based on stochastic reduced-order models \cite{field2015efficacy} to capture the material and processing variability associated with laser welding and subsequent propagation to prediction of component reliability.
Rappel \etal \cite{rappel2018identifying} used Bayesian calibration for elastoplasticity models while accounting for errors in observed stress and strain measurements in addition to additive modeling error.
Ku\v{c}erov\'{a} and Matthies \cite{Kucerova:2010} applied Bayesian updating of a material property of an arbitrary heterogeneous material described by a Gaussian random field relating to a manufactured specimen.

The contributions of this work are multifold.
The primary contribution is that we treat diverse categories of uncertainties related to the resolvable and sub-threshold porosity in plasticity and failure in a unified, physically motivated framework.
For this application we are able to map the observed variability to distributions of physical parameters.
The method we develop could be directly applied to other materials with voids or inclusions, and could be extended to treat materials, \eg polymers and wood, with other microstructural sources of variability .
Particular methodological novelties that enable these results are:
(a) the formulation and application of multivariate normal approximation to the full likelihood for embedded variability calibration (\sref{sec:embedded}),
(b) the development of pseudo-marginal MCMC for dealing with an extremely high-dimensional source of uncertainty derived from a microstructural (porosity) field (\sref{sec:nuisance}),
(c) the use of KLE to model 3-dimensional porosity process (previously only 2-dimensional models were constructed \cite{Ilango:2016}), calibrated with statistical information extracted from CT scan images (\sref{sec:microstructures}),  and
(d) the use of radial basis functions in the construction of effective, accurate surrogate models of strain- and stress-based QoIs of the ensemble of mechanical responses (\sref{sec:surrogate}).

In \sref{sec:experiment}, we provide a brief overview of the experimental data from Boyce \etal \cite{boyce2017extreme} that informs the model.
In \sref{sec:simulation} we give details of the models for simulating plasticity and the failure process using explicit mesh based realizations of the observed porosity and sub-resolution porosity implicit in the constitutive damage model.
In addition, we describe how we represent and reduce the very high dimensional space of porosity to a Karhunen-Lo\`eve  process based on the observed voids.
In \sref{sec:UQ}, we develop the uncertainty quantification methods to treat four classes of uncertain parameters, namely:
(a) the observable porosity,
(b) physical parameters of the constitutive damage model,
(c) the parameters of embedded parametric uncertainty used to represent the variability across the ensemble of samples, and
(d) the noise/discrepancy between the complete porosity-damage model and the data.
Then, in \sref{sec:results}, we discuss the results of the model selection and calibration.
We also draw conclusions about the proportion of the observed variance explained by the explicitly represented porosity and the model-based damage.
Finally, in \sref{sec:conclusion}, we summarize results and discuss their importance to future work.

\section{Experimental Data} \label{sec:experiment}

\begin{figure}
\centering
{\includegraphics[width=0.5\textwidth]{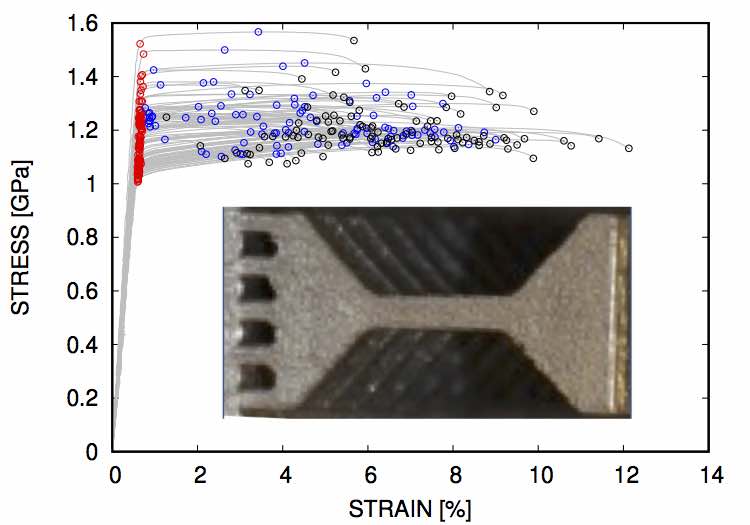}}
\caption{Stress-strain response of dogbone specimens (inset).
Quantities of interest demarked with circles: yield (red), maximum/ultimate stress (blue),  failure strain (black).
}
\label{fig:features}
\end{figure}

\begin{figure}
\centering
{\includegraphics[width=0.65\textwidth]{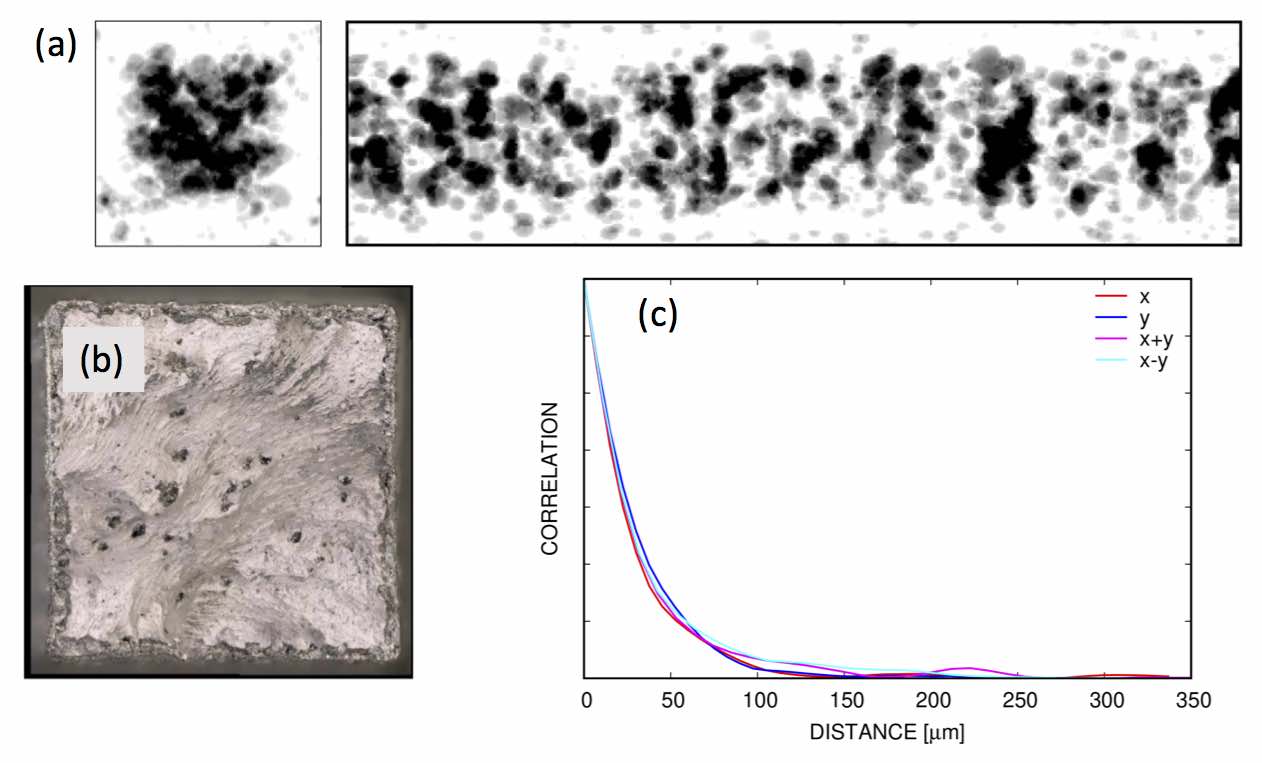}}
\caption{
Porosity of the dogbone specimens (a) core of the gauge region showing through thickness void probabilities (maximum 10\%), (b) post-failure cross-section showing surface ``crust'' and interior computed tomography (CT) scanned region, and (c) spatial correlation in various directions.
The observable mean porosity is 0.008 and spatial correlation length is on the order of 50 $\mu$m.
}
\label{fig:pores}
\end{figure}

The experiments of Boyce \etal \cite{boyce2017extreme} provide tensile stress-strain data quantifying the mechanical response up to failure (refer to \fref{fig:features}) and corresponding computed tomography (CT) scans that reveal the internal porosity of the AM 17-4PH stainless steel dogbone specimens (refer to \fref{fig:pores}).
One build provided 120 nominally identical replica dogbones with 1mm$\times$1mm$\times$4mm gauge sections, a subset of each build were imaged over the entire gauge section with CT prior to mechanical testing.
For this study we examined 105 stress-strain curves and 18 CT scans from the same build.

All the tensile tests were performed at a 10$^{-3}$/s strain rate effected by grips engaging the dovetail ends of the specimen seen in the inset of \fref{fig:features}.
Engineering stress was measured with a load cell, the cross-section measured by CT and optical techniques, and the strain was determined by digital image correlation (DIC) of the gauge section.
\fref{fig:features} illustrates the ensemble of tensile tests which display variation in their elastic response, yield and hardening behavior, and failure characteristics.
Clearly, the variability tends to increase across the ensemble from the elastic to the failure regime.
From these experimental stress-strain curves we extracted a number of QoIs, $\FEATURES$:
the {\it effective} elastic modulus $\bar{E}$ from the initial slope,
the yield strength $\sigma_Y$ from a 0.001 offset strain criterion,
the yield strain $\epsilon_Y$ from  the strain corresponding to the yield stress,
the ultimate tensile strength $\sigma_U$ from the maximum stress,
the ultimate tensile strain $\epsilon_U$ from the strain corresponding to the maximum stress,
the failure strain $\epsilon_f$ from maximum strain achieved, and
the failure stress $\sigma_f$ stress corresponding to maximum strain.
Model calibration described in \sref{sec:UQ} utilized this data, summarized in \tref{tab:features}, with the exception of $\bar{E}$ since it is highly correlated with $\sigma_Y/\epsilon_Y$.
From \fref{fig:features} and \tref{tab:features} it is apparent that there is significant variability in the mechanical response, with strain-based QoIs having greater variance than stress-based QoIs due to the flatness of the post-yield stress curve.

The microstructural source of the variance in mechanical response was partially revealed by the CT of the  0.75mm$\times$0.75mm$\times$4mm interior of the gauge sections of some of the specimens.
The CT scans were performed on each dogbone independently with 7.5 $\mu$m resolution, utilizing the same scan parameters.
Identification of porous voxels was performed by applying a (per sample) threshold to the per-voxel CT intensity, and voids were identified by the spatial connectivity of porous voxels.
The interior region of the samples appear to have a fairly homogeneous distribution of voids as \fref{fig:pores}a indicates.
However, post-fracture analysis such as the cross-section shown in \fref{fig:pores}b have a distinct surface crust which is approximately 0.05mm thick.
Within the central CT scanned region, the void distribution appears to be isotropic, as the spatial correlations averaged across all samples in different directions through the cross-section shown in \fref{fig:pores}c indicate.
The mean porosity across all samples examined for this study was $\meanporosity=$ 0.008 and spatial correlation length was on the order of 50 $\mu$m.
The average porosity varied sample-to-sample $\approx 60\%$ which is reflected in the $\approx 10\%$ variance of the effective elastic modulus $\bar{E}$ reported in \tref{tab:features}.

\begin{table}[t]
\centering
\begin{tabular}{|lc|ccc|}
\hline
Parameter & & Min & Mean$\pm$deviation & Max \\
\hline
Elastic modulus (GPa)              &
$\tilde{E}$             & 165.0  & 218.4 $\pm$ 22.5 (10.3\%) & 277.8  \\
Yield            strength (GPa)    &
$\sigma_{Y}$     & 1.02   &  1.17 $\pm$ 0.11 ( 9.8\%) & 1.54  \\
Yield            strain (\%) &
$\epsilon_{Y}$   & 0.53   &  0.67 $\pm$ 0.08 (11.5\%) & 0.93  \\
Ultimate tensile strength (GPa)    &
$\sigma_{U}$   & 1.10   &  1.23 $\pm$ 0.09 ( 7.0\%) & 1.57  \\
Ultimate tensile strain  (\%)      &
$\epsilon_{U}$ & 0.77   &  4.49 $\pm$ 2.24 (50.0\%) & 9.00  \\
Failure strength (GPa)             &
$\sigma_{f}$     & 1.07   &  1.20 $\pm$ 0.08 ( 6.8\%) & 1.53  \\
Failure strain (\%)                &
$\epsilon_{f}$   & 1.27   &  6.34 $\pm$ 2.14 (33.8\%) & 12.13  \\
\hline
\end{tabular}
\caption{Experimental response quantities of interest $\FEATURES$ from 105 tensile tests of AM 17-4PH stainless steel tensile specimens.
Note that $\tilde{E}$ is the {\it effective} (observed) elastic modulus, which should not be confused with the model parameter $E$ introduced in \sref{sec:BCJ}.
}
\label{tab:features}
\end{table}

\section{Simulation Methods} \label{sec:simulation}

Since the CT scans where only able to resolve voids with diameters larger than the 7.5 $\mu$m pixel size and there is evidence \cite{heckman2019} that a population of sub-threshold pores exist initially and can grow and nucleate, we employed a hybrid modeling approach.
Specifically, we chose to represent the resolvable pores explicitly in finite element meshes and the sub-threshold pores via damage mechanics (an implicit homogenized porosity field governed by an evolution equation).
The selected viscoplastic damage constitutive model is described in detail in \sref{sec:BCJ}.
Coupling between the explicit and implicit voids is provided by an ``element death'' mechanism that allows highly damaged elements to act like voids by no longer contributing to the local stress response and thereby adding to the explicit porosity.
To enable sufficient sampling of the porosity realizations and avoid the added complexity of boundary reconstruction schemes, we chose to voxelate the porosity on structured meshes.
We set the mesh resolution to be comparable to the CT scan resolution; this aspect will be discussed in more detail in \sref{sec:microstructures}.
Furthermore, the CT scans did not cover the entire gauge section nor the distinct crust and are limited in number, so we created a model, detailed in \sref{sec:microstructures}, to provide sufficient statistically similar realizations for the uncertainty quantification discussed in \sref{sec:UQ}.
For efficiency, we only simulated the gauge section with simple tension effected by minimal Dirichlet boundary conditions.

\subsection{Constitutive Model} \label{sec:BCJ}

Given the observed phenomenology and the need to model a population of sub-threshold voids that may nucleate and grow, we selected an isothermal variant of the Bammann-Chiesa-Johnson (BCJ)  viscoplastic damage model \cite{bammann1996modeling,horstemeyer1999void,brown2012validation,karlson2016sandia}.
The Cauchy stress $\sigmab$ is given by a linearly elastic mixture rule, which assumes an additive decomposition of elastic and plastic strain:
\begin{equation} \label{eq:stress}
\sigmab = (1-\phi) \Cbb ( \epsilonb - \epsilonb_p ) \ ,
\end{equation}
based on the void volume fraction $\phi$, where $\Cbb$ is the (isotropic) elastic modulus tensor with components $[ \Cbb ]_{ijkl} = E/(1+\nu) \left( \nu/(1-2\nu) \delta_{ij}\delta_{kl} + 1/2 (\delta_{ik}\delta_{jl} + \delta_{il}\delta_{jk}) \right)$ which depends on Young's modulus $E$ and Poisson's ratio $\nu$.
Here $\epsilonb$ is the total strain, $\epsilonb_p$ is the plastic strain, and $\epsilonb - \epsilonb_p$ is the elastic strain.
The plastic strain ${\epsilonb}_p$ evolves according to:
\begin{equation} \label{eq:dotep}
\dot{\epsilonb}_p = \sqrt{\frac{3}{2}} f \sinh^n \left( \frac{\sigma/(1-\phi) - \kappa}{Y} - 1 \right) \frac{\sb}{\| \sb \|}
\end{equation}
with the isotropic hardening $\kappa$ governed by:
\begin{equation} \label{eq:hardening}
\dot{\kappa} = (H-R \kappa) \, \dot{\epsilon}_p
\end{equation}
where $\kappa(t=0) = \kappa_0$.
Here, $\sb = \dev \sigmab$ is the deviatoric stress, $\sigma \equiv \sqrt{\frac{3}{2} \sb\cdot\sb}$ is the von Mises stress, $Y$ is the yield stress, $f$ is the flow coefficient, $n$ is the flow exponent, $H$ is the hardening modulus, $R$ is the recovery coefficient, and $\dot{\epsilon}_p = \sqrt{2/3 \dot{\epsilonb}_p \cdot \dot{\epsilonb}_p}$ is the equivalent plastic strain rate.
Recovery is associated with dislocation annihilation, recrystallization, and related processes.

The implicit void volume fraction $\phi$ is associated with material damage and is related to void concentration $\eta$ and the average void size $v$ by $\phi \equiv \eta v / ( 1 + \eta v )$.
For nucleation, the void concentration $\eta$ (a number density) evolves according to:
\begin{equation} \label{eq:etadot}
\dot{\eta} = \left( N_1 \left(\frac{2^2}{3^3} - \frac{J_3^2}{J_2^3}\right) + N_3 \frac{p}{\sigma} \right) \eta \, \dot{\epsilon}_p
\end{equation}
where $p = 1/3 \, \sigmab\cdot\Ib$ is the pressure, $J_2 = 1/2 \tr \sb^2 = 1/3 \, \sigma^2$ and $J_3 = 1/3 \tr \sb^3$.
The $N_1$ component responds to torsion, and the $N_3$ component corresponds to nucleation due to tension/compression (triaxiality), refer to Ref. \cite[Table 2]{horstemeyer1999void}.
The growth of the average void size $v$ is given by:
\begin{equation} \label{eq:nudot}
\dot{v} = \sqrt{\frac{2}{3}} \frac{1+\eta v}{\eta} \left( \left(1+\eta v\right)^{m+1} -1\right) \sinh\left(\frac{2 (2 m-1)}{2 m + 1} \frac{p}{\sigma}\right) \ ,
\end{equation}
where $m$ is the damage exponent.
As shown in Karlson \etal \cite{karlson2016sandia} and in Brown and Bammann \cite{brown2012validation}, \eref{eq:nudot} can be extended to include the effect of newly nucleated voids, assuming them to have volume $v_0$:
\begin{equation} \label{eq:nudot1}
\dot{v} = \sqrt{\frac{2}{3}} \frac{1+\eta v}{\eta} \left( \left(1+\eta v\right)^{m+1} -1\right) \sinh\left(\frac{2 (2 m-1)}{2 m + 1} \frac{p}{\sigma}\right) - (v - v_0)\frac{\dot{\eta}}{\eta}.
\end{equation}
Then, \eref{eq:etadot} and \eref{eq:nudot1} can be combined, and using arguments in Brown and Bammann \cite{brown2012validation} in the limit of vanishing volume of newly nucleated voids, the evolution relation for the void volume fraction (porosity), $\phi$ is obtained:
\begin{equation} \label{eq:phidot}
\dot{\phi} = \sqrt{\frac{2}{3}}\dot{\epsilon}_\text{p}\frac{1 - (1-\phi)^{m+1}}{(1-\phi)^{m}}\sinh\left(\frac{2(2m-1)}{2m+1}\frac{p}{\sigma} \right) + (1-\phi)^2 \dot{\eta} v_0.
\end{equation}
Lastly, once the void fraction $\phi$ exceeds a threshold $\phi_\text{max}$ the material, as discretized by the finite element mesh, is considered completely failed.

\subsection{Generation of Synthetic Microstructures} \label{sec:microstructures}

To develop models that are consistent with experimentally-observed failure metrics while accounting for the resolvable porosity, we need a means of generating mesh-based realizations of the CT visible porosity.
Various approaches have previously been proposed in the literature to model randomly porous media \cite{Quiblier:1984,Adler:1990,Yeong:1998}.
We used a Karhunen-Lo\`eve expansion (KLE) (see \eg \cref{le2010spectral}) to model porous media as a random process through an intermediate Gaussian random process.
KLE is a mean-square optimal representation of square-integrable stochastic processes and has been widely used in many engineering and scientific fields.
Much like a Fourier series representation, KLE represents a stochastic process using a linear combination of orthogonal functions; however, KLE differs from Fourier series in that the coefficients are random variables (as opposed to deterministic scalar quantities) and the basis depends on the correlation function of the process being modeled (as opposed to pre-specified harmonic functions).

Ilango \etal \cite{Ilango:2016} proposed a methodology to construct KLE models (see \aref{app:kle} for a synopsis and generalization) for 2-dimensional porosity processes. Herein, we will extend the methodology to construct KLE models for the 3-dimensional porosity process of the core and crust regions such as those visible in \fref{fig:pores}b. Given the experimental data, we assumed that the binary random process modeling porosity $\porosity(\xb)$ is homogeneous and isotropic \cite{Guttorp:1994,Adler:1990}, with a two-point correlation function given by
\begin{equation}\label{disc_process_R}
\text{R}_{\porosity\porosity} \left( \xb_1, \xb_2  \right) = \expectation \left[ \porosity( \xb_1 ) \porosity( \xb_2 ) \right] = \text{R}_{\porosity\porosity}(r)
\end{equation}
where $r = \| \xb_1 - \xb_2 \|$ is the distance between positions $\xb_1$ and $\xb_2$.

A modeling choice was made to construct one KLE model for the ensemble of specimens as opposed to a separate KLE model for each dogbone sample.
This choice was justified by the fact that the extracted empirical correlation function did not vary significantly across samples of the core region (no data was available for the crust), and thus our aim was to represent the variance of the physical response across an ensemble of structures produced under nominally identical conditions.
With this choice, we approximated the statistics: mean porosity $\meanporosity$ and spatial correlation $\text{R}_{\porosity \porosity}$, as ensemble averages utilizing the available CT scans across all samples of the porous media as described in \cref{Adler:1990}.
The ensemble average was taken to be the average over all specimens and all available pairwise combinations of voxels and associated distances.
The constructed KLE representation of $\porosity(\xb)$ was subsequently used to generate realizations by sampling $\porosity(\xb)$ on a structured mesh covering the nominal gauge section of the tensile specimens and at a resolution commensurate with the CT voxels.
Specifically, we generated realizations of these processes on a structured grid of 135$\times$135$\times$534 voxels identical to the CT scan with 7.5$\mu$m pixel resolution, where the outer 10 voxels in a cross-section were designated as the crust and the remainder as core.

Starting with the core region, we obtained a mean porosity of $\meanporosity$= 0.008  and \fref{fig:pores}c provides the experimental correlation function $\text{R}_{\porosity \porosity}(r)$ as a function of distance $r$ between voxels.
There is noise in the underlying experimental data which is predominantly attributed to (a) a finite sample of specimens being analyzed and (b) the process of obtaining and post-processing CT images.
In order to filter out this noise while satisfying physical constraints relating to correlation functions, we fit the data to the widely-used power-exponential correlation function \cite{Ripley:1981,Cressie:1991}:
\begin{equation}\label{eq:power_exp}
\text{R}(r) = \exp \left( - \left( \frac{r}{\kappa} \right)^\rho \right) ,
\end{equation}
where $\kappa > 0$ is the correlation length and $\rho \leq 2$ is a tunable parameter.
Calibration of this correlation function to the experimental data resulted in a correlation length $\kappa = 0.0526 {\rm mm}$ and power $\rho = 1.122$.
In order to capture 99.99\% of the energy (measured in terms of variance) of the process as realized on the structured grid of 115$\times$115$\times$534 voxels corresponding to the core region, the KLE was truncated at 10,676 terms, equivalent to 0.18\% out of 5,887,350 terms (number of voxels)
A similar model was constructed for the crust region.
Lacking CT data of the crust, we assumed the correlation length of its binary porosity process was approximately the same as that for the core region, with a mean porosity ten times of that of the core.
The increased porosity of the crust was surmised from \fref{fig:pores}b and expert knowledge.
This assumption resulted in a truncated KLE for the intermediate Gaussian process with 32,366 terms to capture 99.99\% of the energy of the process (0.33\% of available terms).

This methodology essentially maps a set of known random variables, namely $\{{\vartheta_j}\}_{j=1}^{L}$, being independent standard Gaussian random variables, to the binary random process representing the (per element) random porosity $\porosity_I$ on the finite element mesh (see \aref{app:kle} for details).
In subsequent analysis, we will group this set of random variables into a random vector $\KLE$, possessing a dimensionality of 43,042 in this specific application.

\subsection{Three-dimensional Plasticity-damage Simulations}\label{simulations}

Since damage models, like the selected BCJ model described in \sref{sec:BCJ}, are well known to create mesh dependent results, we needed to tie the model calibration to a fixed element size and  we initially chose the  mesh size to be the same as the CT pixel size ($134\times134\times534$ elements).
A full CT resolution mesh for a single realization required about 72 hours of computation time on 256 cores to simulate a tension test.
Given the number of simulations needed for the uncertainty quantification (UQ) analysis ($\gg$ 10$^2$, see \sref{sec:surrogate}) we could not afford to use full CT resolution meshes.
Instead we systematically applied a coarsening method to the full resolution KL realizations.
We found that structured meshes with $26\times26\times106$ elements lead to an acceptable computational cost of $\approx$ 16 CPU-hours per sample.

To obtain a coarsened version of the full resolution porosity realization we used a threshold to determine whether a coarsened element covering approximately 5$^3$ voxels represented a void or solid material.
A voxel that intersected two or more coarse elements was counted by only one of the coarse elements; and, if the fraction of void voxels within the coarse element was above a threshold, the entire element was set as a void.
This threshold parameter was tuned across an ensemble of 50 full resolution porosity realizations and the threshold value 0.35 best matched the mean and variance of the original realizations (0.00819$\pm$0.0114 for the original  \vs 0.00846$\pm$0.00131 for the coarsened representations).
We checked the mechanical response of one full resolution mesh versus that of its coarsened counterpart.
For both the ultimate tensile strength and the failure strain, the QoI values of the fine and coarse realizations were within 0.3\%.

\subsection{Calibration Parameter Selection}\label{parameter_selection}
In order to reduce the high dimensionality of the tunable constitutive model parameters $\{E,\nu,Y,\ldots\}$, we fixed a number of parameters based on numerical considerations, expert knowledge, and preliminary sensitivity studies.
\tref{tab:parameters} summarizes the division into parameters that were calibrated $\PHYSICAL$ and those that were held fixed.

In making this determination, we took the perspective that the basic elastic-plastic parameters $E$, $\nu$, $Y$ represent the well-determined, intrinsic properties of the fully dense, undamaged 17-4PH material and hence are fixed, pre-determined constants; consequently, we allow other parameters to account for the observed variable response due to microstructure.
For instance, we designated the initial damage $\phi_0$ as a parameter for calibration to account for variance in the observed effective modulus $\tilde{E}$ (refer to \tref{tab:features}).
Likewise, the initial hardening $\kappa_0$ has a confounding effect in determining the observed yield stress $\sigma_Y$ (refer to \eref{eq:dotep}), so it alone was allowed to largely determine this QoI.
Preliminary studies allowed us to set the maximum damage for element death parameter $\phi_\text{max}$ at 0.5 to emulate the failure characteristics of the experimental data; and, we found the model response was not particularly sensitive to this value.
Additional preliminary QoI sensitivity studies using Sobol indices ranked the sensitivity of the torsion nucleation parameter $N_1$ orders of magnitude below competing parameters, whereas the sensitivity of the failure QoIs to the triaxiality nucleation parameter $N_3$ was highly ranked; hence, $N_1$ was fixed and $N_3$ was left for calibration.

To specify the initial density $\eta_0$ and initial size at nucleation, $v_0$, of the unresolved voids we took the value $\eta_0 v_0$ = 10$^{-4}$ for a similar material from \cref{karlson2016sandia}, estimated $v_0$ = 0.1 $\mu$m$^3$ based on the size of the CT voxels (421.5 $\mu$m$^3$), and arrived at $\approx$ 30 voids per finite element.
This population is sufficient to satisfy the basic premise of the damage model, \ie that there is an ensemble of uncorrelated voids present.
The remaining parameters, flow exponent $n$, damage exponent $m$, and flow coefficient $f$ were fixed at published values, see \eg \cref{karlson2016sandia}; preliminary studies showed that the response was relatively insensitive to $f$ as well.
Physical reasoning and accepted values for similar materials were used to center the ranges of the calibration parameters $\PHYSICAL = \{ R, H, \kappa_0, \phi_0, N_3\}$.
Both the hardening $H$ and recovery $R$ parameters (refer to \eref{eq:hardening}) were included in this set to capture the range of post yield slopes and maximum stresses observed in the data.
The ranges of the calibration parameters given in \tref{tab:parameters} were obtained through an iterative process involving the bounds of parameter space explored by the Monte Carlo sampler described in \sref{sec:UQ}.

\begin{table}[t]
\centering
\begin{tabular}{|lc|c|}
\hline
Parameter & & Value \\
\hline
Young's modulus (GPa)         & $E$       & 240 \\
Poisson's ratio               & $v$       & 0.27 \\
Yield strength (MPa)          & $Y$       & 600   \\
Initial void size ($\mu$m$^3$)    & $v_0$   & 0.1 \\
Initial void density ($\mu$m$^{-3}$) & $\eta_0$  & 10$^{-3}$  \\
Flow exponent                 & $n$        & 10 \\
Damage exponent               & $m$        & 2 \\
Torsion nucleation            & $N_1$      & 10      \\
Flow coefficient              & $f$        & 10        \\
Maximum damage                & $\phi_\text{max}$  & 0.5  \\
\hline
Isotropic dynamic recovery    & $R$        & [0.1, 10]\\
Isotropic hardening (GPa)     & $H$        & [2, 10]        \\
Initial hardening (MPa)       & $\kappa_0$ & [200, 600]    \\
Initial damage                & $\phi_0$   & [0.0001, 0.2] \\
Triaxiality nucleation        & $N_3$      & [5, 15]       \\
\hline
\end{tabular}
\caption{Parameter values for fixed parameters (upper) and ranges for calibration parameters (lower) $\PHYSICAL = \{ R, H, \kappa_0, \phi_0, N_3 \}$.
}
\label{tab:parameters}
\end{table}

\subsection{Response Surrogate Model} \label{sec:surrogate}

To calibrate likely values for the parameters $\PHYSICAL$ given the experimental data, we used Bayesian model calibration techniques developed in \sref{sec:UQ}.
This statistical inversion relies on the joint sampling of the posterior PDF of the parameters $\PHYSICAL$ using Markov chain Monte Carlo (MCMC) sampling-based strategies.
The joint characterization of the posterior parameter PDF of 5 independent parameters usually requires at least $10^6$ samples (at 16 CPU-hours per sample), where each sample is a forward model simulation of the computationally-intensive finite element model described in \sref{simulations}.
One popular strategy that reduces this computational burden relies on surrogate models of the physical response.
Such surrogates capture the complex, often nonlinear, mapping from the unknown parameters $\PHYSICAL$ to observable, system QoIs $\FEATURES$.
Although not general purpose, QoI surrogate models are designed to be relatively cheap to evaluate while being of sufficient accuracy.

\begin{figure}[h!]
\centering
\includegraphics[width=0.95\textwidth]{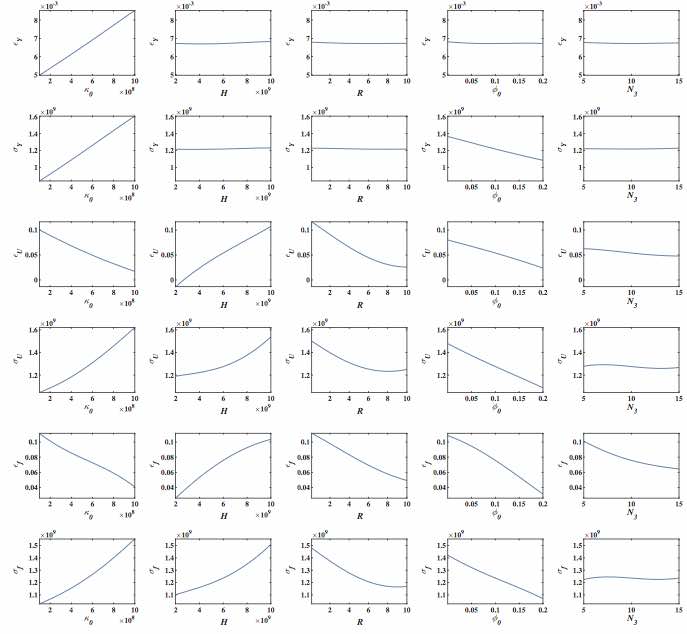}
\caption{One-dimensional response surfaces for each of the six quantities of interest.
For each plot, the other four parameters are fixed at the nominal (mid-range) values.
}
\label{fig:resp_surf}
\end{figure}

Following the strategy used in in a previous study \cite{rizzi2019bayesian}, we first attempted to employ a polynomial chaos expansion (PCE)~\cite{Wiener:1938,Ghanem:1991} to construct surrogates over the domain of plausible values of the unknown parameters.
Such a methodology exploits a global polynomial basis for capturing nonlinear input-output mappings (as exhibited by the response surfaces for the observed QoIs in \fref{fig:resp_surf}).
Specifically, PCE surrogates require higher-order bases to capture the nonlinearities accurately. This implies an exponential growth in the number of unknown PCE coefficients to tune, which is proportional to the number of forward model simulations required to construct such surrogates.
For this work, we first attempted the construction of third-order PCE surrogates in the 5-dimensional parameter space.
The PCE coefficients were obtained using Galerkin projection utilizing 241 simulations corresponding to quadrature points obtained using Smolyak’s sparse tensorization and the nested Clenshaw-Curtis quadrature formula \cite{le2010spectral}. The accuracy, as measured using normalized root-mean-square error with 100 additional simulations corresponding to Monte Carlo samples, was deemed insufficient for subsequent calibration tasks. Such surrogates also exhibited slow convergence against increasing PCE order from 1 through 3. In such situations, one may attempt to construct higher order PCE surrogates, but a 4th order PCE surrogate would have required a prohibitive 801 sparse-quadrature points (simulations) for construction. Given the unacceptable computational cost of pursuing this approach, we decided to abandon the PCE-based approach, and to turn to an approach using radial basis functions (RBF's) \cite{Hardy:1971,Dyn:1986,Powell:1987}.

Specifically, we used RBF's to construct surrogate response models for each physical QoI $\FEATURES$ as a function of the calibration parameters $\PHYSICAL$, since they are one of the most capable multidimensional approximation methods \cite{Jin:2001}.
For each porosity realization constructed using the methodology outlined in \sref{sec:microstructures}, we fit a RBF surrogate for each of the 6 QoIs $\FEATURES$ reported in \tref{tab:features}.
One could construct a global surrogate (over all realizations) per QoI to cover the space of KLE coefficients used in modeling the explicit porosity process as well as the unknown parameters, but we opted against that approach due to the extremely high-dimensionality of the KLE coefficient space (see \sref{sec:microstructures} for details).
Instead, we constructed surrogate models for the observed QoIs in terms of only the unknown parameters over a finite number of porosity realizations (one surrogate per QoI for a specific porosity realization), with the number of porosity realizations ($<$ 100) being significantly lower than the dimensionality of the porosity process ($>$ 10,000).
To construct such approximations, we started with a set of parameter points $\PHYSICAL_i$, $i = 1, \hdots, n$, and corresponding QoI values, $\feature_i = \feature\left( \PHYSICAL_i \right)$ for each QoI $\feature_i$ in $\FEATURES$.

The RBF approximation, $\feature \approx \tilde{\feature} \left( \PHYSICAL \right)$, is a linear combination of basis functions (kernels) that depend on the distances between the evaluation point, $\PHYSICAL$, and a set of kernel centers, $\PHYSICAL_J$, $J = 1, \hdots, M$ given by
\begin{equation}
\tilde{\feature} \left( \PHYSICAL \right) = \sum_{J=1}^M c_J \mathcal{K} \left( \left\Vert \PHYSICAL - \PHYSICAL_J \right\Vert \right)
\end{equation}
where $\left\Vert \cdot \right\Vert$ denotes the Euclidean norm, and coefficients $c_J$ are to be determined.
Widely-used kernels include Gaussian, $\mathcal{K} \left( r \right) = e^{-\left( \varepsilon r \right)^2}$, multiquadratic, $\mathcal{K} \left( r \right) = \sqrt{1+\left( \varepsilon r \right)^2}$, and cubic, $\mathcal{K} \left( r \right) = \varepsilon r^3$, where $\varepsilon$ is a tunable shape parameter (see \cref{Gutmann:2001} for a more complete list of RBF kernels).
After preliminary comparative investigations (results omitted for brevity), we chose to utilize Gaussian kernels, since they lead to sufficiently-low cross-validation errors in this context.
Regarding the choice of shape parameter $\varepsilon$, there exist many rules of thumb as well as data-driven approaches (see \cref{Fasshauer:2007} for a brief survey).
For simplicity, we chose to use the value of $D/\sqrt{M}$ as recommended by Franke \cite{Franke:1982}, with $D$ representing the diameter of the smallest hypersphere containing all data points.
Since we work with normalized parameters in the range of $[-1,1]$, $D$ is then equal to $2 \sqrt{n_p}$, with $n_p=5$ being the dimensionality of the parameter space $\PHYSICAL$.
A quick sensitivity study yielded no significant increase in accuracy of the resulting surrogates with varying choice of $\varepsilon$ about this recommended value.

In many applications, the centers $\{ \PHYSICAL_J \}$ are chosen to be the data points $\{ \PHYSICAL_J \} = \{ \PHYSICAL_i \}$, with $M = n$.
The constants $c_J$, in that setting, may be determined by having the approximation exactly match the given data at the data points, leading to an interpolative approximation function based on collocation.
In this realistic application, the QoIs extracted from the simulations were corrupted by relatively small, yet significant, noise relating to errors in the QoI-extraction from stress-strain curves sampled at a finite set of strain values.
In other words, the true value $f_i = \tilde{\feature} \left( \PHYSICAL_i \right)$, corresponding to parameter realization $\PHYSICAL_i$, was known only up to a small amount of unknown noise.
In this case, an interpolative RBF approximation is equivalent to fitting the unknown noise along with the underlying QoI.
More generally, over-fitting leads to erroneous surrogates even in the noise-free case due to parameter identifiability issues.

The issue of over-fitting the noise in the extracted QoIs may be alleviated through regularization \cite{Wendland:2005}, for example.
We reduced the effect of noise by choosing the number of centers to be less than the number of available data points, \ie $M < n$, as form of projection.
An advantage of this choice is the reduced cost in evaluating the trained RBF surrogates.
The specific number of centers $M$ is a modeling choice that was made using cross-validation, with the leave-one-out cross-validation (LOOCV) error available in closed-form \cite{Arlot:2010}.
The locations of both sets of data points and centers for RBF surrogate construction were obtained using Latin hypercube sampling (LHS) \cite{Sacks:1989}, resulting in a sample set that was predominantly random, but uniform in each separate dimension.
Although more optimal strategies exist for choosing the data points and centers adaptively (see \cref{Wei:2012} for example), we chose this space-filling design.

\begin{figure}[h!]
\centering
\includegraphics[width=0.84\textwidth]{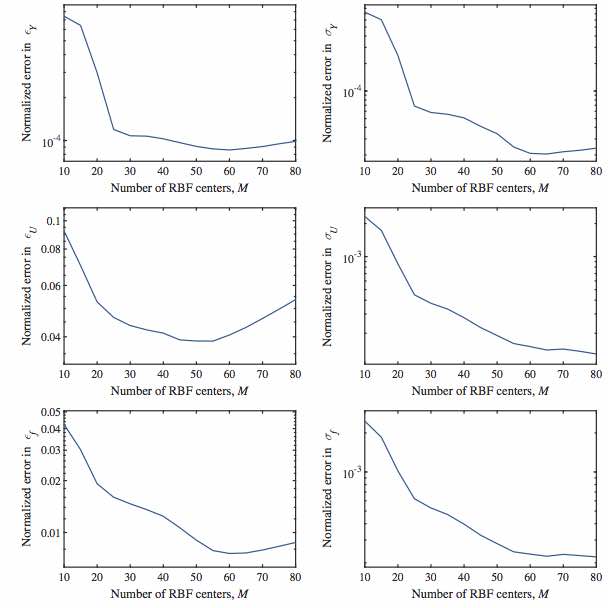}
\caption{Normalized leave-one-out cross-validation error in the RBF surrogates with varying number of RBF centers for the observed quantities of interest}
\label{fig:rbf_loocv}
\end{figure}

To obtain data at points $\{ \PHYSICAL_i \}$ for each porosity realization, we ran $n=200$ forward model simulations at a set of corresponding LHS samples in parameter space with parameter ranges given in \tref{tab:parameters}.
The extracted QoIs $\FEATURES_i$ from those simulations were used in training the RBF surrogates.
We varied the number of RBF centers $\PHYSICAL_J$, $J = 1,\dots M$ using $M$ from 20 to 80, and extracted the LOOCV error for each trained RBF surrogate.
To obtain the average LOOCV error, normalized by the corresponding QoI, the process was repeated over the 50 available porosity realizations, and a further 30 times (per porosity realization) over different LHS realizations of the centers.
The results are shown in \fref{fig:rbf_loocv}.
Overall, a choice of $M = 50$ centers results in RBF surrogates with close to minimal error for all QoIs, an increase in $M$ results in negligible reduction in the LOOCV error for some QoIs and a substantial increase in error for the noisier QoIs $\epsilon_U$ and $\epsilon_f$.

As an illustration of the mapping between system inputs $\PHYSICAL$ and outputs $\FEATURES$, \fref{fig:resp_surf} shows the response surfaces (in the form of RBF surrogates) for each of the six QoIs $\FEATURES$ as a function of one of the five parameters $\PHYSICAL$, while fixing all other parameters at the nominal (mid-range) values.
From this set of plots, it is clear which parameters are influencing which QoIs.
For instance, yield QoIs $\epsilon_Y$ and $\sigma_Y$ are largely determined by the initial hardening $\kappa_0$ and damage $\phi_0$, as the constitutive model described in \sref{sec:BCJ} and physical intuition would suggest.

\section{Bayesian Calibration} \label{sec:UQ}

In this section, we develop a combined treatment of the aleatory uncertainties related to the explicit porosity configurations and the epistemic uncertainties of the material parameters.
The procedure involves an ensemble of mesh-based realizations of the observable porosity $\porosity_I$ (described in \sref{sec:microstructures}) and marginalization over the resulting response surrogates.
We separate the parameters into three categories:
(a) the high dimensional porosity as represented by the realizations controlled by the KLE parameters $\KLE$,
(b) the physical parameters of interest including the ``embedded'' error/Beta distributions for a subset of the parameters $\PHYSICAL$, and
(c) the hyper parameters for the additive uncorrelated measurement noise intensities $\HYPER$.
The corresponding {\it stochastic} measurement model of the experimental data $\DATA$ is
\begin{equation} \label{eq:model}
d_{i}(\PHYSICAL,\KLE,\hyper_i) = \feature_i(\PHYSICAL;\KLE) + \hyper_i \noise_i
\end{equation}
where $\feature_i$ is the model predicted QoI given synthetic porosity realization (parametrized by KLE coefficients $\KLE$) and $\eta_i$ is the corresponding independent additive, zero-mean, unit-variance, Gaussian noise, scaled by parameter $\hyper_i$. We can rewrite \eref{eq:model} in vector format, grouping all observed QoIs and model-predicted counterparts into vectors $\bd $ and $\FEATURES$, respectively, to obtain
\begin{equation} \label{eq:model_vectorized}
\bd (\PHYSICAL,\KLE,\HYPER) = \FEATURES(\PHYSICAL ;\KLE) + \HYPER \NOISE \ ,
\end{equation}
with $\HYPER$ denoting the diagonal matrix with the diagonal elements $\HYPER_{i,i} = \hyper_i$, and $\NOISE$ being the vector of identical independently distributed (IID) zero-mean, unit-variance, Gaussian random variables $\noise_i$.
The available data $\DATA$ is comprised of the six QoIs $\bd^{(j)} = \FEATURES^{(i)}$ (see \tref{tab:features}) as extracted from 105 tensile tests ($j = 1,\hdots,105$), one per available AM dogbone specimen.

\subsection{Inference in the Presence of High-Dimensional Aleatory Uncertainty: Pseudo-Marginal Likelihood Approach} \label{sec:nuisance}

Generally speaking, the statistical calibration of models using experimental data involves inferring a set of unknown, or weakly known, parameters.
Often, a subset of those parameters are not directly relevant for subsequent analysis (such as sensitivity analysis or forward propagation of uncertainty).
Such parameters, generally termed ``nuisance'' parameters, are still important in the calibration process and are thus jointly inferred with the parameters of interest.
The parameters $\hyper_i$ scaling the measurement noise $\noise_i$ are a common example of this kind of parameter.

In this section, we describe how we deal with a different kind of nuisance parameter vector, $\KLE$, which is used to model (via Karhunen-Lo\`eve expansion described in \sref{sec:microstructures}) the high-dimensional random process.
This is the observable porosity field $\porosity(\xb)$, which acts as an uncontrollable source of uncertainty in the physical system, \ie the location and sizes of voids.
In UQ terms, such a source of uncertainty contributes to {\it aleatory} uncertainty \cite{Bedford:2001}, arising from the inherent variability of a phenomenon (in this case the random porosity distribution) which cannot be further reduced from the available data.
Here, this means that we cannot rely on knowing the locations of voids when making predictions.
In contrast, the unknown parameter vector $\PHYSICAL$ of interest contributes to epistemic uncertainty \cite{Bedford:2001} arising from incomplete knowledge of the phenomenon and can be reduced with the available data $\DATA$.
In this context, both types of uncertainty have probabilistic characterization.
However, modelers utilizing UQ techniques make this distinction as the two types of uncertainties require separate treatment in a Bayesian setting (see \cref{Winkler:1996} for a philosophical discussion on the need to separate sources of uncertainties in this fashion).

In such a setting, one performs joint inference of the uncertain parameters and nuisance parameters.
The joint posterior PDF of the uncertain parameter vector $\PHYSICAL$ and nuisance parameter vector $\KLE$ is first decomposed using Bayes' law:
\begin{equation}\label{joint1}
p\left( \PHYSICAL, \KLE \,\vert\, \DATA \right) \propto p\left( \DATA \, \vert \, \PHYSICAL, \, \KLE \right) p\left( \PHYSICAL, \, \KLE \right) \ ,
\end{equation}
with the first and second right-hand-side terms corresponding to the joint likelihood and prior PDFs of the vectors $\PHYSICAL$ and $\KLE$, respectively.
We proceed by inferring the marginal posterior PDF of $\PHYSICAL$, instead of joint inference of $\PHYSICAL$ along with $\KLE$.
Since $\KLE$ is of high dimensionality (more than 43,000 dimensions) in comparison to $\PHYSICAL$ (5-dimensional), we marginalize over $\KLE$ using a few Monte Carlo realizations (50 samples) distributed according to the marginal prior PDF, $p\left( \KLE \right)$, of $\KLE$.
This allows us to construct surrogate models for the observed QoIs in terms of $\PHYSICAL$ over a limited number of realizations of $\KLE$, and thus significantly reduce the computational cost involved in forward model simulations.
The other justification for this marginalization-based approach is the lack of requirement for a joint uncertain characterization of $\PHYSICAL$ and $\KLE$ since, as described, $\KLE$ parametrizes an uncontrollable source of uncertainty in the physical system, \ie the location and sizes of voids across different AM specimens, and, hence, is not a quantity of direct engineering design interest.
Note that for subsequent uncertainty propagation analysis, we will utilize a marginal posterior PDF of $\PHYSICAL$ derived with this approach.

Towards that end, we first rewrite \eref{joint1} as
\begin{equation}\label{joint2}
p\left( \PHYSICAL, \KLE \,\vert\, \DATA \right) \propto p\left( \DATA \, \vert \, \PHYSICAL, \, \KLE \right) p\left( \PHYSICAL \right) p\left( \KLE \right) \ ,
\end{equation}
which assumes $\PHYSICAL$ and $\KLE$ are independent prior to the assimilation of data.
(Given the lack of previous analysis/knowledge for a joint characterization this is a reasonable assumption to be made in practice.)
Consequently, the marginal posterior PDF of $\PHYSICAL$ is
\begin{equation}\label{marginalized_posterior1}
p\left( \PHYSICAL \,\vert\, \DATA \right) \propto \left[ \int p\left( \DATA \, \vert \, \PHYSICAL, \, \KLE \right) p\left( \KLE \right) {\rm
d}\KLE \right] p\left( \PHYSICAL \right) \ .
\end{equation}
The integral in \eref{marginalized_posterior1} corresponds to the marginal likelihood of $\PHYSICAL$, being the likelihood of $\PHYSICAL$ averaged over $\KLE$, weighted by the prior PDF of $\KLE$, $p\left( \KLE \right)$.
In general, this marginal likelihood PDF does not have an analytical solution and one must resort to deterministic or stochastic schemes of numerical integration.
One conceptually simple approach utilizes Monte Carlo (stochastic) integration to arrive at the following approximation $\widehat{p} \left( \PHYSICAL \,\vert\, \DATA \right)$ for the marginal posterior:
\begin{align}\label{marginalized_posterior2}
\widehat{p} \left( \PHYSICAL \,\vert\, \DATA \right) & \propto \left[ \frac{1}{N_{_\text{MC}}} \sum_{k = 1}^{N_{_\text{MC}}} p\left( \DATA \,\vert\, \PHYSICAL , \KLE_k \right) \right] p\left( \PHYSICAL \right) \ ,
\end{align}
with $p\left( \DATA \,\vert\, \PHYSICAL , \KLE_k \right)$ being the likelihood of the parameter vector $\PHYSICAL$ conditional on the specific realization of nuisance parameter vector $\KLE_k$ and those realizations are distributed according to the prior PDF $p\left( \KLE \right)$.
For subsequent sampling of $\PHYSICAL$ according to the approximate posterior PDF in \eref{marginalized_posterior2}, $\widehat{p} \left( \PHYSICAL \,\vert\, \DATA \right)$, we rely on adaptive Metropolis-Hastings Markov chain Monte Carlo (MCMC) sampling  \cite{gamerman2006markov,berg2008markov}.
In this context, the Monte Carlo approximation to the marginal likelihood falls under a general class of Pseudo-marginal Metropolis–Hastings methods \cite{Andrieu:2009}, with our specific approach being a variation of the Monte Carlo within Metropolis (MCWM) algorithm \cite{ONeill:2000}.
More specifically, our variant of MCWM (suggested but not pursued in \cref{Andrieu:2009}) relies on the same set of realizations of $\KLE$ to compute the marginal likelihood for all samples of $\PHYSICAL$ as generated by the MCMC simulation.
This allows us to construct and reuse surrogates over a fixed ensemble of $\KLE$, namely $\KLE_k$, $k = 1, \hdots, N_{_\text{MC}}$, corresponding to KLE realizations of the porosity process (obtained using Monte Carlo sampling of KLE coefficients).

\begin{figure}[h!]
\centering
\includegraphics[width=0.95\textwidth]{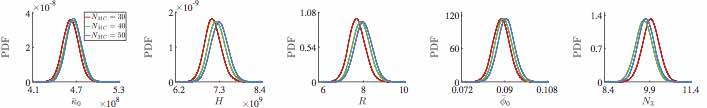}
\caption{Parameter PDFs. Convergence with increasing number of realizations of the porosity for marginal likelihood computation.
}
\label{fig:pseudo_marginal_conv}
\end{figure}

In the next section, we will fully investigate the convergence properties of this pseudo-marginal likelihood approach using a simple, computationally inexpensive elasticity problem.
Subsequently, we will show convergence results relating to the marginal posterior PDFs of the five parameters of interest for the costly, porosity/failure problem with increasing number of Monte Carlo samples $N_{_\text{MC}}$ in \fref{fig:pseudo_marginal_conv}.
These results were obtained using the methodology already discussed and methods to handle the epistemic aspects as detailed in \sref{sec:embedded}.
Increasing the number of Monte Carlo samples of the porosity process for approximating the marginal likelihood from 30 to 40, and finally to 50 generated posterior PDFs that were relatively close to one another, suggesting that 50 realizations of the porosity were sufficient for engineering analysis of this problem.

\subsection{Demonstrative Mechanics Example of the Pseudo-Marginal Likelihood Approach} \label{sec:embedded:example}

To analyze the convergence properties of the Pseudo-Marginal Likelihood approach in a tractable way, we consider the following governing stochastic partial differential equation (SPDE) describing the one-dimensional axial deformation of a rod \cite{Clough:1975} with stochastic geometry
\begin{align}
\frac{ \partial }{ \partial x } \left[ E A \left( x; \KLE \right) \frac{ \partial u }{ \partial x } \right] + f \left( x \right) & = 0 \ , \ \ \ \ \ \ \ \ \ \ x \in \Omega \coloneqq \left( 0, 1 \right) \nonumber \\
u \left( 0 \right) = u \left( 1 \right) & = 0 \ ,\label{eq:bvp} \\
f \left( x \right) & = 1 \ , \nonumber
\end{align}
with $u$ being the displacement due to deformation, $f$ being the axially-applied distributed force, $E$ being the elastic modulus, and $A$ being the stochastic, spatially-varying cross-sectional area (inducing heterogeneity) determined by the vector of KLE coefficients $\KLE$ (introduced in \sref{sec:microstructures}).

\begin{figure}
\centering
{\includegraphics[width=0.5\textwidth]{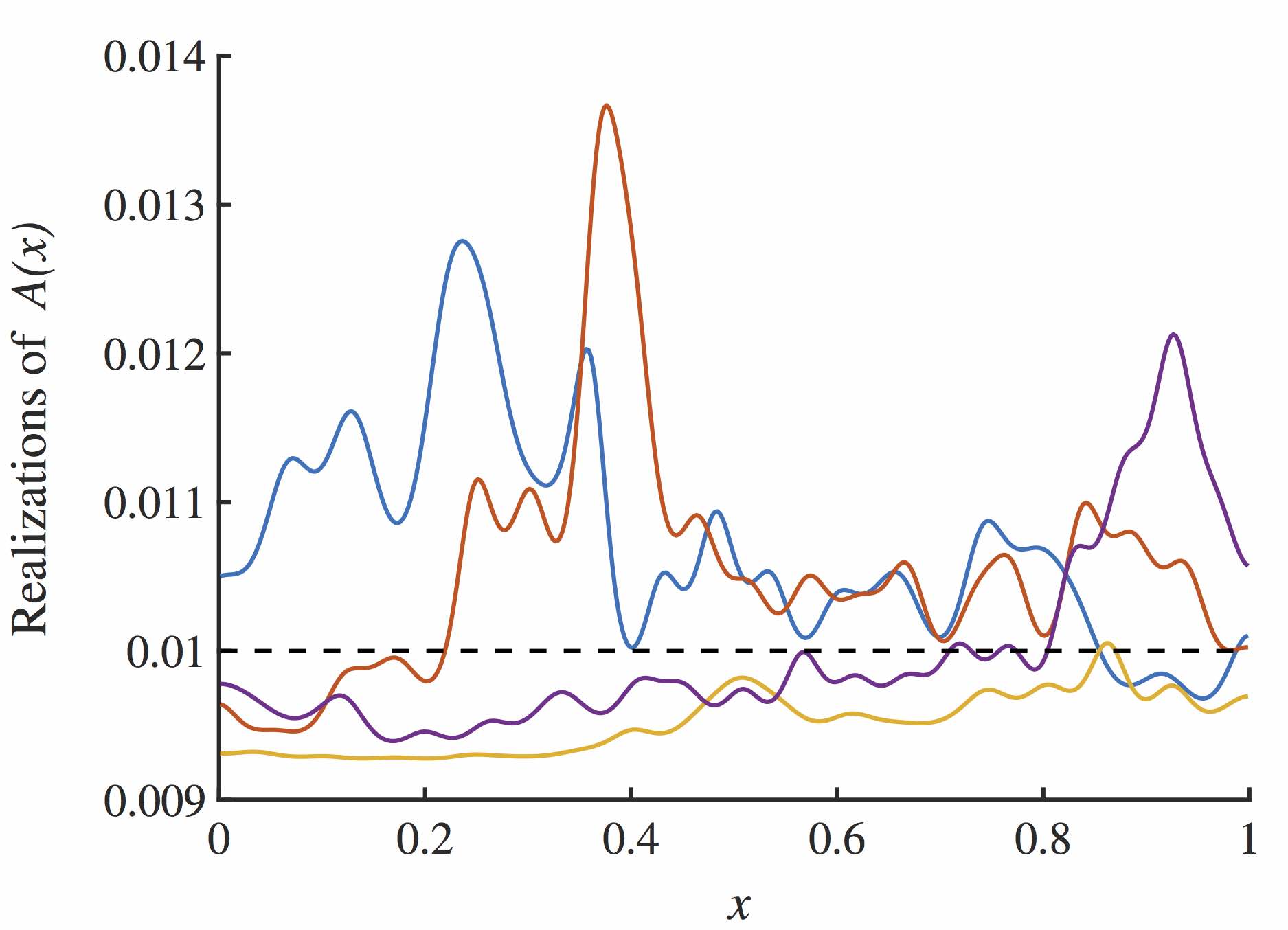}}
\caption{Axial deformation problem: Monte Carlo realizations of the stochastic cross-sectional area process, $A \left( x, \KLE \right)$, with coefficient of variation $\epsilon = 0.1$ and cross-correlation length $b=0.5$. The dashed line represents the mean process.}
\label{fig:toy:A_reals}
\end{figure}

Treating this as an inverse problem, we take the parameter $E$ to be an unknown quantity to be inferred from a set of 10 noisy observations of the mid-span deflection (\ie $u(x=1/2)$, one per different random realization of $A(x)$ (\ie a total of 10 repeated experiments).
The observational noise is assumed to be independent and identically distributed (IID) Gaussian with zero mean and variance of $5 \times 10^{-7}$ (in comparison to an average mid-span deflection of $1.25 \times 10^{-5}$).
We will discretize \eref{eq:bvp} using the finite-element method (FEM) with 200 linear finite elements, which proves sufficient resolution to represent the KLE modes.
The stochastic cross-sectional area $A(x; \KLE)$ is modeled as a log-normal random process, \ie
\begin{equation}
A \left( x; \KLE \right) = \bar{A} \left[ 1+ \epsilon \alpha\left(x, \KLE \right) \right] \ ,
\end{equation}
where $\bar{A} = 0.01$ is the mean cross-sectional area, $\epsilon$ is the coefficient of variation, and $\alpha$ is a log-normal process:
\begin{equation}
\alpha \left( x, \KLE \right) = \frac{1}{\sqrt{e\left(e-1 \right)}} \exp{ \left( g \left( x, \KLE \right) - e^{ \frac{1}{2}} \right)}  \ ,
\end{equation}
with zero mean and unit variance.
Here, $g \left( x, \KLE \right)$ is a zero-mean, unit-variance Gaussian stochastic process with an exponential covariance function \cite{le2010spectral}:
\begin{equation}
{\rm E} \left[ g \left(x, \cdot \right) g \left(x', \cdot \right) \right]  = \exp{ \left( - \frac{\left\vert x - x' \right\vert}{b} \right)} \ ,
\end{equation}
where $b$ is the correlation length of the underlying Gaussian random process.
Such choice of covariance function is convenient in this setting as it leads to analytical expressions for the eigenvalues and eigenmodes in Karhunen-Lo\`eve (KL) representations (see \cref{le2010spectral}).
The available KLE representation simplifies the generation of Monte Carlo realizations of the log-normal cross-sectional area (samples thereof are shown in \fref{fig:toy:A_reals} for $b=0.5$ and $\epsilon=0.05$), acting as an uncontrollable source of uncertainty in the physical system (similar to the effect of random porosity).
Since we are dealing with a domain of unit length, the correlation length $b$ can be considered to be a dimensionless quantity (\ie normalized by the range of $x \in [0,1]$).

\begin{figure}
\centering
{\includegraphics[width=0.5\textwidth]{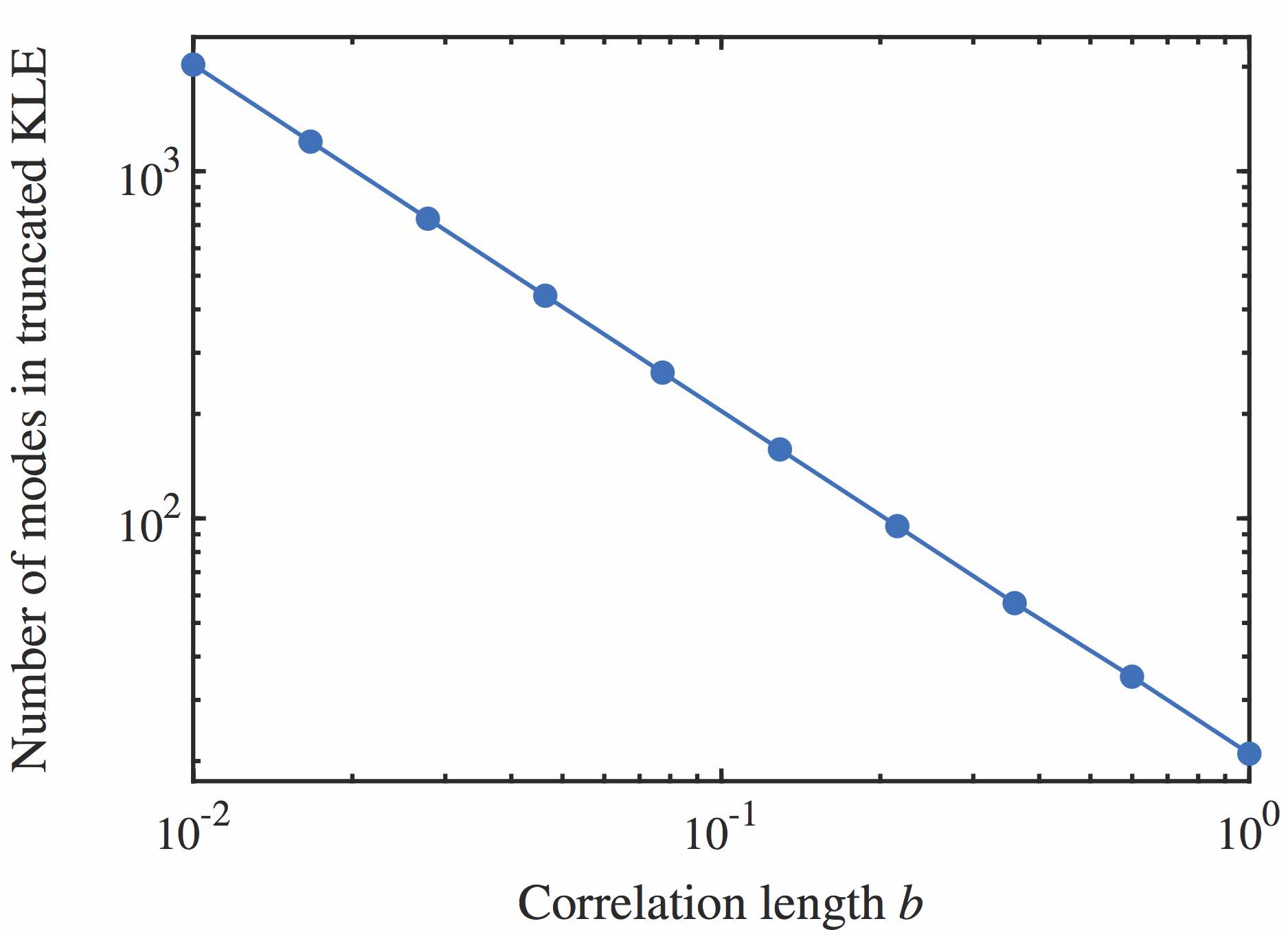}}
\caption{Axial deformation problem: the correlation length of the underlying Gaussian random process for the cross-sectional area vs the number of eigenvalues of the corresponding truncated KL representation required to capture 99\% of the energy of the process.
}
\label{fig:toy:n_eigs}
\end{figure}

\begin{figure}
\centering
{\includegraphics[width=0.5\textwidth]{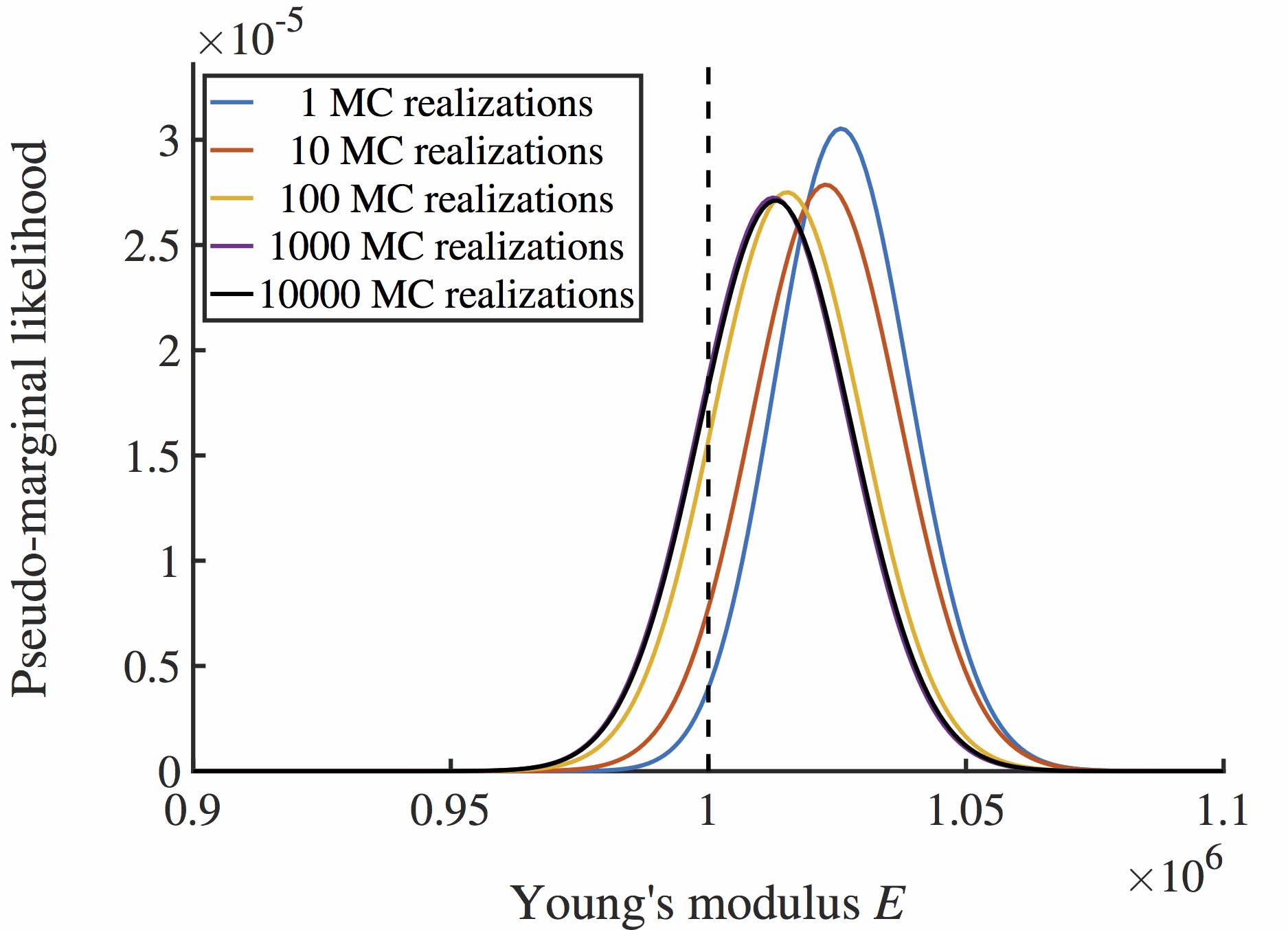}}
\caption{Axial deformation problem: pseudo-marginal likelihood PDF of Young's modulus, $E$, with a cross-sectional area process having coefficient of variation $\epsilon = 0.1$ and cross-correlation length $b=0.5$.
Distributions for increasing number of Monte Carlo realizations of the stochastic cross-sectional area process, $A \left( x; \KLE \right)$, used in the marginalization process.
The dashed vertical line marks the true value of $E$.}
\label{fig:toy:E_lik}
\end{figure}

For the range of correlation lengths considered in this study, $b \in \left( 0.01, 1 \right)$, we construct corresponding KLE representations of $g \left( x, \KLE \right)$, truncated to capture 99\% of the energy (variance) as measured by the accumulated sum of eigenvalues.
The relationship between the correlation length and the corresponding number of eigenvalues of the truncated KL representation is shown in \fref{fig:toy:n_eigs} and displays near perfect power-law behavior.
It is obvious that the dimensionality of the cross-sectional area process, ranging between 20 and 2000, far exceeds that of the parameter vector of interest (in this case comprising of merely one parameter, $E$).
Therefore, we utilize the marginal likelihood approach outlined in the previous section to obtain samples of the cross-sectional area process, with which we arrive at a Monte-Carlo estimate of the marginal likelihood for parameter $E$.
We vary the number of Monte Carlo realizations of $A$ used in estimating the marginal likelihood in \eref{marginalized_posterior2} to study the convergence of the pseudo-marginal likelihood for Young's modulus, $E$.
Using a cross-sectional area process with a coefficient of variation $\epsilon = 0.1$ and underlying cross-correlation length $b=0.5$, we obtain the pseudo-marginal likelihoods shown in \fref{fig:toy:E_lik} with increasing number of cross-sectional realizations.
In this case, we see rapid convergence of the likelihood, with nearly Gaussian characteristics.

It is evident that both mean and mode estimates of Young's modulus are greater than the true value in this example.
Over/under-estimation of inferred parameters in Bayesian calibration exercises with finite and noisy data sets is to be expected.
Upon close examination, the ten available noisy data points relating to the midspan deflection are 30\% smaller than the mean midspan deflection, as normalized by the standard deviation of the midspan deflection.
This is due to the specific realizations of the random noise that contaminates the observations.
Such under-representation of the scatter in the midspan deflection results in over-estimation of Young's modulus, \ie an overall increase in system stiffness (due to greater $E$ values) better explains the lower observed values of the midspan deflection.

\begin{figure}
\centering
{\includegraphics[width=0.5\textwidth]{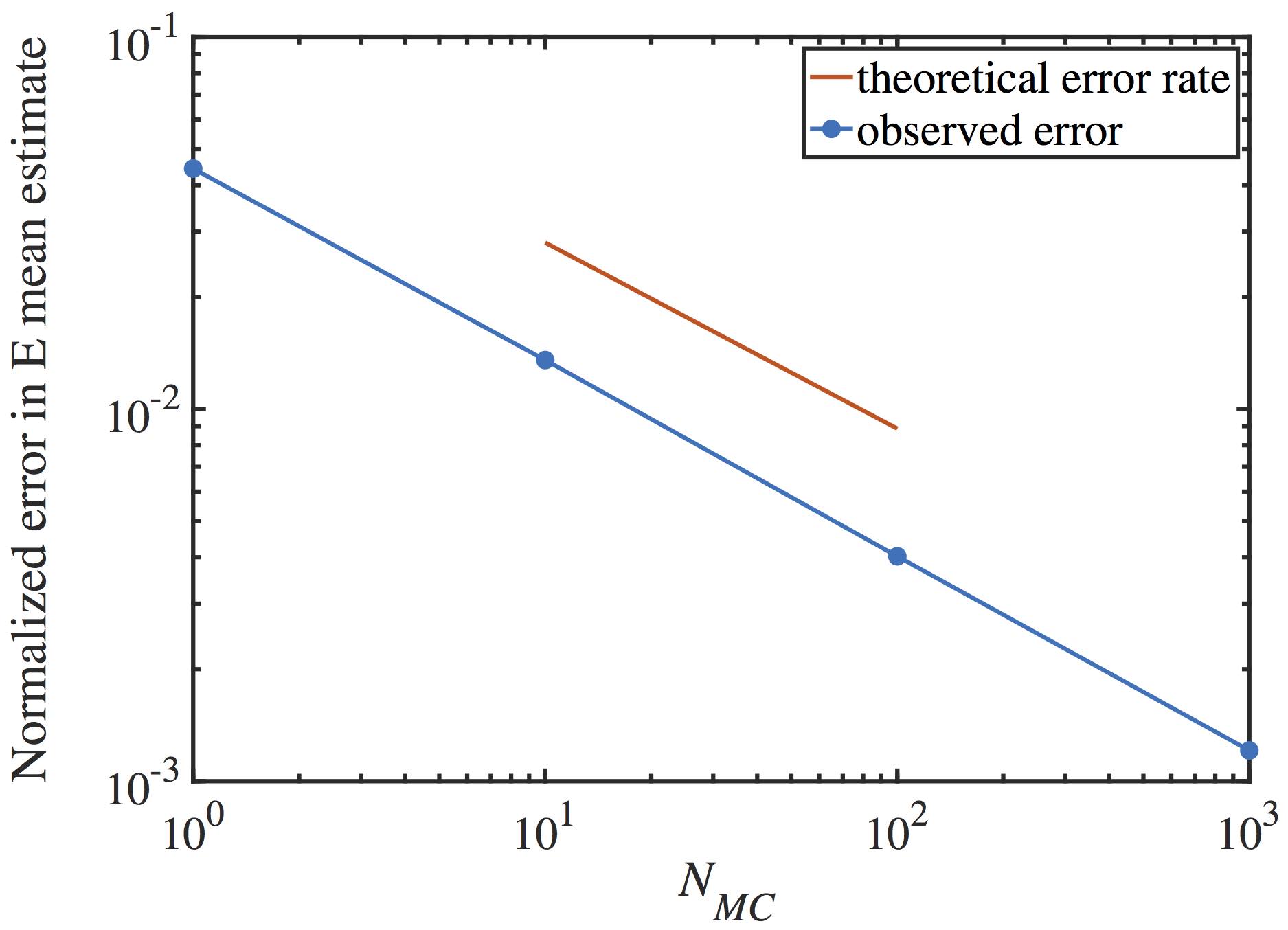}}
\caption{Axial deformation problem: Monte Carlo error (obtained from repeated studies) in extracted mean moment from the pseudo-marginal likelihood with increasing number of Monte Carlo realizations of the cross-sectional area, with a cross-sectional area process having coefficient of variation $\epsilon = 0.1$ and cross-correlation length $b=0.5$.
The error shown (blue curve) is normalized by the true value of $E$.
Also shown is the theoretical error (up to a proportionality constant) for the Monte Carlo method (red trend line).
}
\label{fig:toy:marg_lik_mean_conv}
\end{figure}

In Bayesian calibration studies, one might be interested in posterior moments relating to a parameter of interest.
In this setting, we will examine the error in estimating Young's modulus $E$ using the mean obtained from the pseudo-marginal likelihood PDF with increasing number of realizations of the area $A$.
For a given number of cross-sectional realizations, we repeat the estimation task $10^4$ times to extract the error (measured as the ensemble standard deviation) in the estimated mean.
(The number of repetitions was chosen using a parametric convergence study.)
This error is then normalized by the true value of $E$ and plotted in \fref{fig:toy:marg_lik_mean_conv}.
Also shown is the theoretical error trend for the Monte Carlo method, being proportional to the inverse of the square-root of the number of Monte Carlo realizations used.
The slopes of the two curves match well (the theoretical trend is only determined up to a proportionality constant), except for small discrepancy which may be attributed to a number of factors, including sampling error (finite number of repetitions) and the fact that the theoretical error is valid for large number of Monte Carlo realizations (due to strong reliance on the central limit theorem).
If we are interested in the mean estimate of $E$, we observe that just 10 to 20 realizations of the cross-sectional area process are sufficient in providing a mean estimate of $E$ with a relative error of 1\%.

\begin{figure}
\centering
{\includegraphics[width=0.5\textwidth]{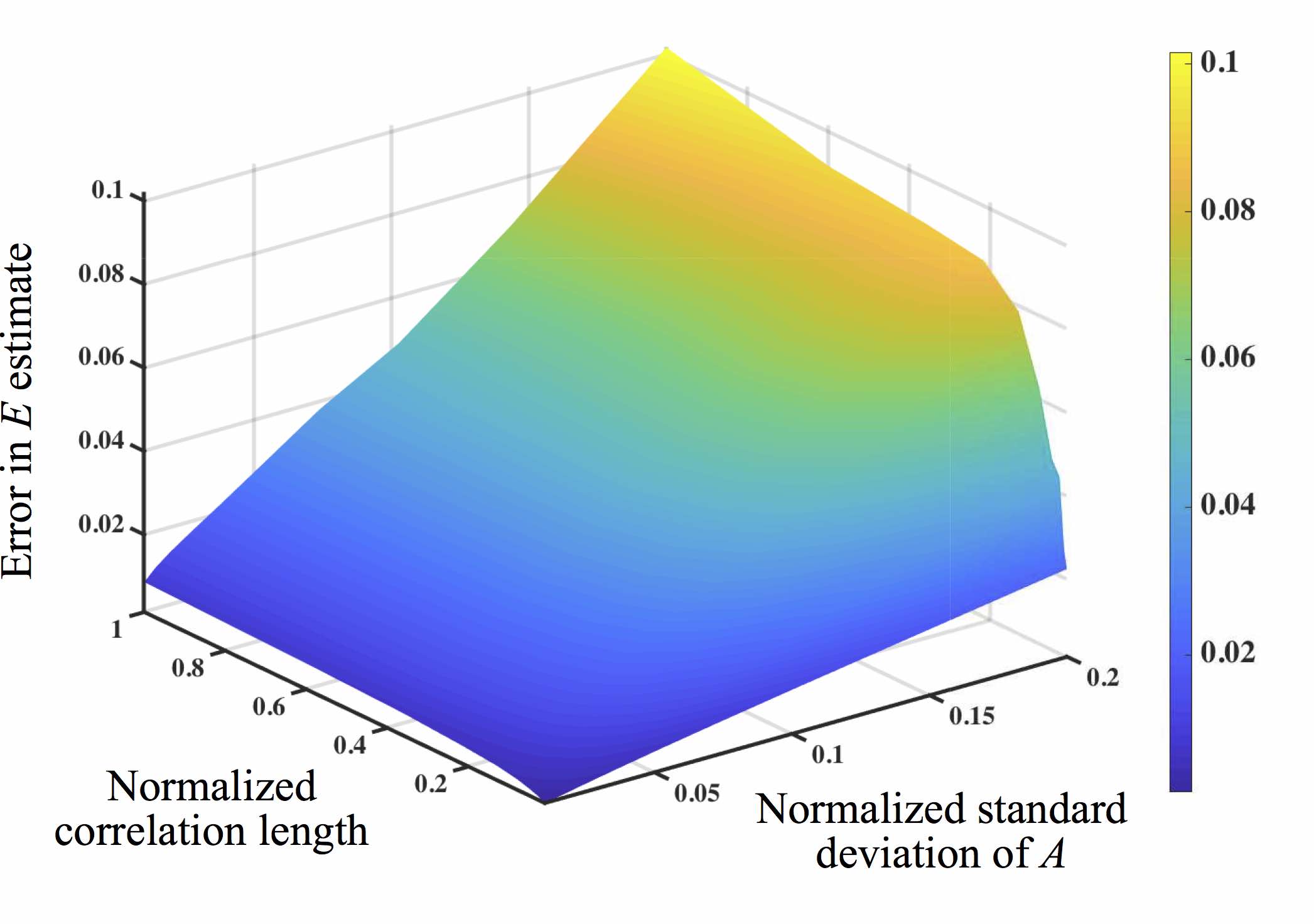}}
\caption{Axial deformation problem: the resulting Monte Carlo error multiplier for the posterior mean estimate of $E$ due to Monte Carlo sampling error in estimating the marginal likelihood as a function of normalized correlation length of the underlying Gaussian random process for the cross-sectional area as well as the uncertainty (normalized standard deviation) in the cross-sectional area random process.}
\label{fig:toy:mc_error}
\end{figure}

Since the rate of convergence of the error in estimating $E$ can be sufficiently captured by the theoretical Monte Carlo error model, we are able to extract the constant of proportionality for that error model for coefficient of variation $\epsilon$ and underlying cross-correlation length $b$ of the cross-sectional area.
That constant, which we will denote as the Monte Carlo error multiplier, is shown in  \fref{fig:toy:mc_error}.
This surface can be used in selecting the number of Monte Carlo realizations of cross-sectional area needed for the pseudo-marginal likelihood to estimate Young's modulus to within certain error.
We see a near-linear dependence on the normalized standard deviation (or coefficient of variation) of the area, in contrast with a more complex dependence on the correlation length.
For small correlation lengths, the error drops quite rapidly.
In the limit of zero correlation length, \ie a white noise process for the area, it seems that a single realization is sufficient (on average) to capture the marginal likelihood for $E$.
This observation helps justify the relatively small number of porosity realizations used in approximating the marginal likelihood for the porosity problem of interest.

\subsection{Embedded Variability} \label{sec:embedded}

In previous sections, we outlined an approach for (a) characterizing the {stochastic} porosity process using the Karhunen-Lo\`eve expansion (\sref{sec:microstructures}), and (b) the subsequent propagation of such a source of variability using Monte Carlo integration in model calibration involving constitutive model parameters $\PHYSICAL$ (\sref{sec:nuisance}).
As will be shown in the following Results section, the explicit porosity variability alone, when propagated through the calibrated model, explains little of the observed variability in QoIs relating to plasticity and failure of AM metals.
In addition to the variability induced by the explicit porosity model, we will embed variability in the constitutive model parameters directly.
Following the methodology first proposed in \cref{sargsyan2015statistical}, certain parameters will be treated as random variables (as opposed to unknown deterministic quantities), and calibrated such that the predicted variability in the mechanical response closely resembles that observed in experiments.
As will be shown in \sref{sec:results}, such embedding of variability in model parameters ameliorates the severe under-representation of variability induced by explicit porosity variability alone, resulting in more consistent scatter of macroscopic performance metrics.
Furthermore, it enables us to associate the distributions of the subset of material parameters with the physical variability of the ensemble of test samples.

We start by representing an unknown constitutive model parameter, $\theta_i$, as a random variable using the following parametrization
\begin{equation} \label{eq:pcPerturbed}
\theta_i = \bar{\theta}_i+ \delta_{\theta_i} \xi_i \ ,
\end{equation}
where $\bar{\theta}_i$ represents the mean coefficient and $\delta_{\theta_i}$ is a scaling parameter that controls the level of variability of $\theta_i$.
Here, $\xi_i$ is a zero-mean independent random variable with a corresponding known PDF, chosen to be one of the commonly used distributions in the family of generalized polynomial chaos (gPC) \cite{Xiu:2002c}.
Such parametrization facilitates the characterization and propagation of uncertainty through computational models.
Popular choices of distributions for continuous random variables include Gaussian, uniform and Beta.
Due to positivity constraints on the model parameters, we would like for the underlying random variables $\xi_i$ to have finite support.
As such, we chose to utilize random variables with Beta PDFs centered at zero rather than uniform due to a modeling preference for gradually decaying PDFs.
The shape parameters for the Beta distribution are chosen to be $a=b=5$, resulting in symmetric and  unimodal density with near-Gaussian shape (but of finite support).
With such choices, \eref{eq:pcPerturbed} is a first-order Jacobi-Beta PCE, with coefficients $\bar{\theta}_i$ and $\delta_{\theta_i}$ to be inferred from the available observations.

Such embedding of variability, once calibrated with one set of observed QoIs, results in a model that allows for the propagation of variability through numerical simulations to any other QoI.
Although we chose first order, independent PCE representations of the parameters $\PHYSICAL$ in \eref{eq:pcPerturbed}, one could utilize higher-order PCEs while simultaneously capturing correlations across model parameters; however, this may drastically increase the number of unknowns (coefficients) to be calibrated.
Another modeling decision relates to the choice of whether to embed variability in each parameter $\theta_i$ or to treat it as an unknown deterministic quantity (equivalent to setting the corresponding $\delta_{\theta_i}$ coefficient to zero and excluding it from the calibration exercise).
A Bayesian model selection technique was used previously \cite{Hakim:2018, rizzi2019bayesian} to determine the optimal embedding of parameter variability, requiring the estimation of Bayes model evidence for each proposed model (with different embedding of variability).
In this investigation, we avoid the costly computation of model evidence and instead rely on selecting a model that achieves a level of scatter of the selected QoIs consistent with observations.
We investigate embedding of the variability in at most two model parameters in $\PHYSICAL$, with a total of 15 different embeddings given the 5 physical calibration parameters.

We proceed with Bayesian inference of the coefficients $\bar{\theta}_i$ and $\delta_{\theta_i}$ grouped into an unknown vector $\EMBEDDED$.
The model of the experimental data in \eref{eq:model_vectorized} with $\PHYSICAL$ physical, $\KLE$ KLE, and $\HYPER$ hyper parameters is now extended to include the embedded variability in the extended physical parameters, $\EMBEDDED$,  as in
\begin{equation} \label{eq:model_emb_error_vectorized}
\bd (\EMBEDDED,\KLE,\HYPER) = \FEATURES(\bar{\PHYSICAL}+ \DELTA \, \xib ;\KLE) + \HYPER \NOISE \ ,
\end{equation}
with $\bar{\PHYSICAL}$ representing the mean vector of the physical parameters, $\DELTA$ the diagonal matrix with the diagonal elements $\DELTA_{i,i}$ being either $\delta_{\theta_i}$ or zero (depending on choice of variability embedding), $\IIDGERMS$ is the vector of IID Beta centered random variables $\xi_i$.
In this context, the embedded variability in $\PHYSICAL$ as well as in the explicit porosity model (parametrized by $\KLE$) are propagated through the model $\tilde{\FEATURES}$ in an attempt to capture the observed variability in the QoIs.
In addition, the additive noise vector $\NOISE$ (scaled by $\HYPER$) aims to capture (a) measurement noise in the recorded stress-strain curves, (b) noise relating to QoI-extraction from stress-strain curves sampled at a finite set of strain values, and (c) any residual variability not captured by the explicit porosity and embedded variability, among other sources of error.
We proceed by inferring the parameters in $\EMBEDDED = \{ \bar{\PHYSICAL}, \DELTA \}$, in addition to parameters in $\HYPER$.
Note that in the absence of embedded parameter variability, $\DELTA \equiv \mathbf{0}$ and $\EMBEDDED$ reduces to the parameter vector $\PHYSICAL$ (with mean coefficients $\bar{\PHYSICAL}$ representing the unknown parameters).

Instead of jointly inferring the explicit porosity, as parametrized by the high-dimensional vector $\KLE$, we followed the methodology outlined in \sref{sec:nuisance} to marginalize over the porosity process, resulting in the following marginal posterior PDF (based on \eref{marginalized_posterior2}) for the parameters of interest:
\begin{align}
\prob(\EMBEDDED , \bhyper \,\vert\, \DATA) & \propto \prob(\DATA  \,\vert\, \EMBEDDED , \bhyper) \, \prob(\EMBEDDED) \, \prob(\bhyper) \nonumber\\
& =  \left[ \frac{1}{N_{_\text{MC}}} \sum_{k = 1}^{N_{_\text{MC}}} p \left( \DATA \,\vert\, \EMBEDDED , \bhyper , \KLE_k \right) \right]  \prob(\EMBEDDED) \, \prob(\bhyper) \ . \label{eq:bayes_embedded}
\end{align}
where $\prob(\EMBEDDED)$ and $\prob(\bhyper)$ are the prior PDFs of the physical and the hyper/noise parameters, respectively; $\prob(\EMBEDDED  , \bhyper \,\vert\, \DATA) $ is the posterior PDF; and $p \left( \DATA \,\vert\, \EMBEDDED , \bhyper , \KLE_k \right)$ is the likelihood function, {\it conditional} on a synthetic realization of the porosity process as parameterized by the KLE coefficients $\KLE_k$.
Recall $\bhyper_i \equiv \HYPER_{i,i}$.
Although this likelihood can be formulated in a number of ways (see \cref{sargsyan2015statistical} for details, and \crefs{Hakim:2018, rizzi2019bayesian} for applications involving other likelihood formulations), we chose to utilize a Gaussian approximation to the full (conditional) likelihood resulting in
\begin{equation} \label{eq:cond_like}
\prob(\DATA  \,\vert\, \EMBEDDED , \bhyper , \KLE_k) = \prod_{j=1}^{N_{\rm tests}} \mathcal{N} \left( \bd^{(j)} ; \boldsymbol{\mu} \left( \EMBEDDED , \KLE_k \right), \boldsymbol{\Sigma} \left( \EMBEDDED , \bhyper , \KLE_k \right) \right) \ ,
\end{equation}
where
\begin{equation} \label{eq:mean}
\boldsymbol{\mu} \left( \EMBEDDED , \KLE_k \right) =
\mathbb{E}_{\xib}[\FEATURES(\bar{\PHYSICAL}+ \DELTA \, \xib ;\KLE_k)]
\end{equation}
and
\begin{equation} \label{eq:std}
\boldsymbol{\Sigma} \left( \EMBEDDED , \bhyper , \KLE_k \right) =
\mathbb{V}_{\xib}[\FEATURES(\bar{\PHYSICAL}+ \DELTA \, \xib ;\KLE_k)] + \HYPER^2
\end{equation}
are the mean vector and covariance matrix of the model predicted QoI vector (6-dimensional in this case) for a given porosity realization $\KLE_k$.
Here, $\bd^{(j)}$ is the observed QoI vector for the $j$-th tensile test (total of $N_{\rm tests} = 105$).
Note that this formulation relies on the assumption of independence between tensile tests, but inter-dependence between the QoIs within each test.
This choice of likelihood formulation is preferable in our context since (a) it avoids non-parametric density estimation (such as kernel density estimation \cite{Silverman:1986}) required by the full or marginal likelihood formulations, (b) it does not rely on Approximate Bayesian Computation (ABC) methods that only measure the discrepancy between a chosen set of statistics of model outputs and the corresponding estimates from the data (rather than a full joint PDF), while (c) it aims to capture joint correlation across parameters (using the covariance matrix) that is informed by the available data.
Although Sargsyan \etal \cite{sargsyan2015statistical} suggested the use of ABC-based formulations due to possible degeneracy and/or deficiencies for certain problems, we encountered no such issues in this context due to the regularization effect of the additive noise term in \eref{eq:model_emb_error_vectorized} as well as the marginalization over the explicit porosity.

For efficiency, the moments in \eref{eq:mean} and \eref{eq:std} were computed using full tensorization of the 1-dimensional, 7-point Gauss–Jacobi quadrature rule.
With these in hand, the evaluation of the conditional likelihood, \eref{eq:cond_like}, follows directly.
In this investigation, we limited the number of non-zero $\delta_{\theta_i}$ to at most two, resulting in at most 49 quadrature points needed for the propagation of uncertainty due to embedded variability through the computational model.
We utilized up to 50 realizations of the porosity process for the marginalization step (\ie $N_\text{MC} \leq 50$), with those realizations fixed throughout the Bayesian calibration exercise (see \sref{sec:nuisance} for justification).
To further expedite the process of computing the pseudo-marginal likelihood in \eref{eq:bayes_embedded}, we utilized RBF surrogates (as described in \sref{sec:surrogate}) constructed for each QoI and porosity realization, with a total of 300 surrogates used in this study.

\section{Results} \label{sec:results}

The application of the Bayesian calibration method described in the previous section results in (a) parameter estimates that we can interpret physically, and (b) response predictions that we can compare to the experimental data.
Recall we have measurement noise levels $\HYPER$ for each of the 6 QoIs of the experimental stress-strain curves $\FEATURES = \{ \epsilon_Y, \sigma_Y, \epsilon_U, \sigma_U, \epsilon_f, \sigma_f\}$, and 5 physical parameters $\PHYSICAL = \{ R, H, {\kappa_0}, {\phi_0}, N_3 \} $.
The calibration also involves a selection of which physical parameters to imbue with embedded variability, \ie possessing distributions not due to epistemic uncertainty, that we assume are representative of the microstructural variability. The posterior PDF is sampled using adaptive Metropolis–Hastings Markov chain Monte Carlo sampling techniques \cite{gamerman2006markov,berg2008markov}, generating a total of $2\times10^5$ samples, $2\times10^4$ of which were discarded as burn-in samples.

\begin{figure}[h!]
\centering
\includegraphics[width=0.85\textwidth]{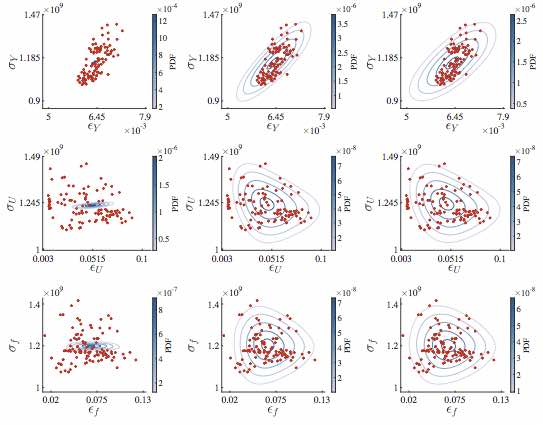}
\caption{Comparison of predictions (contours) with data (red points).
Left column: explicit porosity only; middle column: explicit porosity plus embedded parameter variability.
Right column: explicit porosity and parameter variability plus ``measurement noise''}
\label{fig:predict_kappa0_phi0}
\end{figure}

\begin{figure}[h!]
\centering
\includegraphics[width=0.75\textwidth]{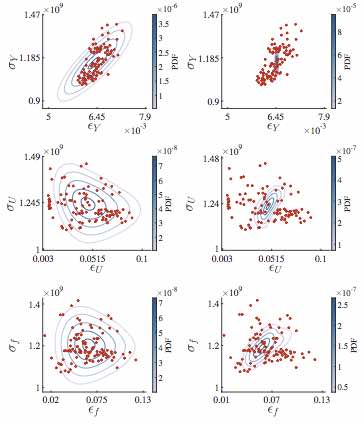}
\caption{Comparison of predictions.
Left column: explicit porosity plus embedded parameter variability in $\kappa_0$ and $\phi_0$.
Right column: explicit porosity plus embedded parameter variability in $\phi_0$ and $N_3$}
\label{fig:predict_compare}
\end{figure}

First we examine the predictions of the response QoIs $\FEATURES$.
\fref{fig:predict_kappa0_phi0} compares variability in yield point (top row), ultimate strength location (middle row) and failure point (bottom row) produced by the ensemble explicit porosity realizations using the maximum \aposteriori (MAP) estimate of the physical parameters $\PHYSICAL$ (left column), the response variability due to the explicit porosity and the embedded parametric variability (center column), and from these two sources plus the calibrated measurement noise (right column).
Overall, the explicit, CT visible porosity alone explains little of the observed variability in these QoIs (refer to \fref{fig:features}); however, its contribution, relative to the spread in the data, does increase as the deformation process progresses from yield to failure.
\fref{fig:predict_compare} compares the optimal choice of embedding variability  in the initial hardening $\kappa_0$ and sub-threshold initial porosity density $\phi_0$ used to generate \fref{fig:predict_kappa0_phi0}, with one of the less explanatory choice of $\phi_0$ and the nucleation parameter $N_3$.
As we can see, the calibrated variability in $\{ \phi_0, N_3\}$ covers less of the data spread in the observed QoIs.
The most apparent discrepancy between the choices is in capturing the scatter in yield strain $\epsilon_Y$, where the calibrated variability in $\{ \kappa_0, \phi_0\}$ outperforms that in $\{ \phi_0, N_3\}$.
Overall, the same can be said for all five other QoIs, albeit to a lesser extent (clearly the spread in $\sigma_U$ and $\sigma_f$ best match the data for this model).
In this setting, the nucleation parameter $N_3$ plays little role in capturing the variability in the QoIs.
This observation is expected if one considers the lack of sensitivity of the QoIs with respect to $N_3$, as is partially illustrated in the one-dimensional response surfaces in \fref{fig:resp_surf}.
Furthermore, the shown sub-optimal choice of embedding variability in $\{ \phi_0, N_3\}$ appears to affect the predicted strains more so than corresponding stresses.
Other combinations of embedding variability in single and pairs of parameters produced similar, sub-optimal results, which are not shown for brevity.

In a previous investigation \cite{rizzi2019bayesian}, we utilized Bayesian model selection to determine the optimal embedding of parameter variability.
In that situation, there were models that exhibited similar level of fidelity in capturing the observed scatter in the QoIs.
Thus we had to rely on information-theoretic approaches in selecting an optimal model.
In this investigation, we found that models apart from the model which embedded variability in the initial hardening and damage, $\{ \kappa_0, \phi_0 \}$, were unable to sufficiently capture the variability in scatter (as measured using maximum likelihood values).
This was true across of the possible permutations of embedding variability in one or two of the five selected physical parameters.
Furthermore, models with embeddings in three or more parameters did not exhibit significant performance gains (again as measured using maximum likelihood values) to warrant the added complexity.
In summary, the model with embedding in $\{ \kappa_0, \phi_0 \}$ is clearly the simplest model that captures the scatter in QoIs.
We therefore proceed with this choice without having to resort to the computationally intensive task of computing the Bayesian model evidence for model selection.

\begin{figure}[h!]
\centering
\includegraphics[width=0.96\textwidth]{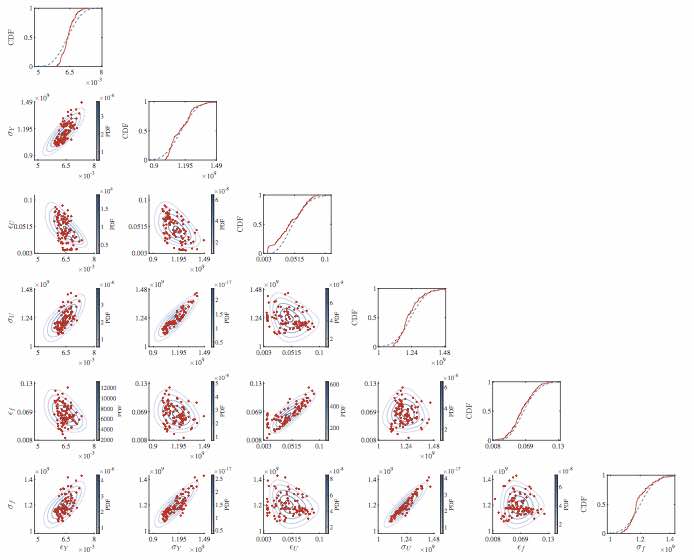}
\caption{Comparison of model predictions and experiments pushing forward variability in explicit porosity and the $\kappa_0$ and $\phi_0$ parameters.
Pairwise joint distributions (off-diagonal) and cumulative distributions (on-diagonal).
}
\label{fig:pdfs1d}
\end{figure}

\fref{fig:pdfs1d} shows the pair-wise joint distributions of the predictions of the six QoIs ($\epsilon_Y$: yield strain, $\sigma_Y$: yield stress, $\epsilon_U$: ultimate strain, $\sigma_U$: ultimate stress, $\epsilon_f$: failure strain, $\sigma_f$: failure stress) and cumulative distribution functions (CDFs) of the individual QoIs compared to data.
Significant positive correlations appear in all the pairwise combinations of the stress QoIs, as well as the ultimate and failure strains, which is in part due to due continuity of the stress-strain curves and the fact they generally do not cross (refer to \fref{fig:features}).
This seems to imply that some specimens are weaker in yield, ultimate and failure strength than others.
The calibrated model represents these physical effects well, as it does the range and general shape of the cumulative density functions (CDFs) constructed empirically from the experimental data.
However, there are some outliers not well captured by the model, particularly a cluster of samples with low ultimate stresses shown in the middle row of \fref{fig:pdfs1d}.

In order to make these predictions, meta parameters ($\HYPER$, 6 parameters, one noise level for each of the response QoIs $\FEATURES$) and physical parameters ($\EMBEDDED$, 7 parameters including the embedded parameters: $\EMBEDDED = \{ \bar{R}, \bar{H}, \bar{\kappa_0}, \bar{\phi_0}, \bar{N}_3, \delta_{\kappa_0}, \delta_{\phi_0} \} $) needed to be calibrated using Bayesian inference.
For the mean physical parameters $\{ \bar{R}, \bar{H}, \bar{\kappa_0}, \bar{\phi_0}, \bar{N}_3  \} $, we chose positive uniform priors \cite{Bolstad:2016} with bounds provided in \tref{tab:parameters}.
For the embedded parameters $\{ \delta_{\kappa_0}, \delta_{\phi_0} \} $, we chose priors that are conditional on the corresponding mean coefficients $\{ \mu_{\kappa_0}, \mu_{\phi_0} \} $ such that the support of the resulting beta distribution for the corresponding physical parameters $\{ \kappa_0, \phi_0 \} $ is within the range given in \tref{tab:parameters}.
Lastly, for the meta parameters $\HYPER = \{ \gamma_{\epsilon_Y}, \gamma_{\sigma_Y}, \gamma_{\epsilon_{U}}, \gamma_{\sigma_{U}}, \gamma_{\epsilon_f}, \gamma_{\sigma_f} \}$ describing the measurement noise intensity in the 6 observed QoIs, we chose to infer the natural logarithm of those meta parameters in order to enforce positivity, with a corresponding uniform prior on [-20,20].

\begin{figure}[h!]
\centering
\includegraphics[width=0.84\textwidth]{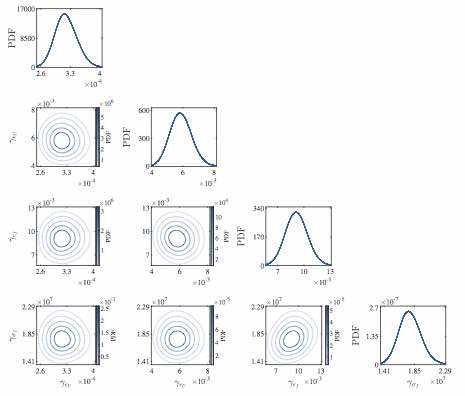}
\caption{Parameter PDFs for the noise hyperparameters $\HYPER$ for each of the response quantities of interest.
Pairwise joint distributions (off-diagonal) and marginalized PDFs for individual parameters (on-diagonal).
Note the estimates of the noise levels $\gamma_{\sigma_Y}$ and $\gamma_{\sigma_U}$ converged to zero and are not shown.
}
\label{fig:pdfs_hyper}
\end{figure}

\begin{figure}[h!]
\centering
\includegraphics[width=0.98\textwidth]{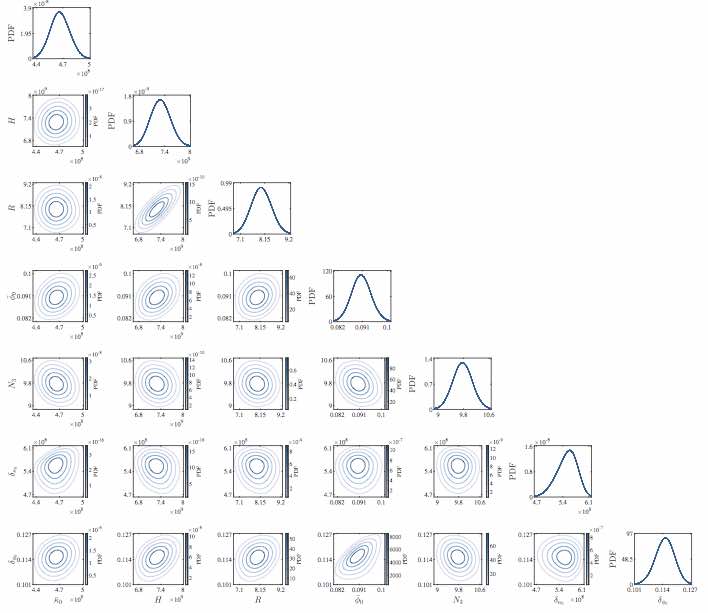}
\caption{Parameter PDFs for the physical parameters $\PHYSICAL = \{ \bar{\kappa}_0, H, R, \bar{\phi}_0, N_3, \delta_{\kappa_0}, \delta_{\phi_0} \}$.
Marginalized PDFs for individual parameters are shown on the diagonal, and pairwise joint distributions are shown on the off-diagonal.
}
\label{fig:pdfs_det}
\end{figure}

\fref{fig:pdfs_hyper} shows the posterior distributions of the noise levels $\HYPER$.
These meta parameters appear to be uncorrelated with bell-shaped marginal distributions.
The noise in the yield stress $\gamma_{\sigma_Y}$ and ultimate stress $\gamma_{\sigma_U}$ converged to negligible values (thus not included in these plots) and the noise in the yield strain $\gamma_{\epsilon_Y}$ and failure stress $\gamma_{\sigma_f}$ are almost negligible; however, the noise in the remaining QoIs, the ultimate strain $\gamma_{\epsilon_U}$ and failure strain $\gamma_{\epsilon_f}$ remains significant.
We associate these values primarily with the inexact QoI extraction, since measurement noise is not apparent in the underlying experimental stress-strain data.
Generally speaking, the physical parameters determining the mean response shown in \fref{fig:pdfs_det} are similarly uncorrelated with bell-shaped, single mode marginal distributions with the exception of the recovery $R$ and hardening $H$ parameters.
Examining the model, particularly the hardening rule \eref{eq:hardening}, this correlation can be seen as the coordinated effect these parameters have in determining the ultimate strength QoIs.
In fact, the ratio $H/R$ determines the ultimate increase in stress beyond yield.

\begin{figure}[h!]
\centering
\includegraphics[width=0.85\textwidth]{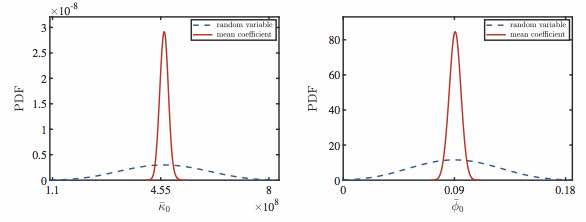}
\caption{Parameter PDFs for the embedded parameters, $\kappa_0$ (initial hardening) and $\phi_0$ (initial damage), characterizing microstructural variability.
Both the uncertainty in the mean (red) and the full distribution of the parameter (blue) are shown.
}
\label{fig:pdfs_emb}
\end{figure}

Finally, the distributions of the embedded parameters shown in \fref{fig:pdfs_emb} indicate the significant amount of microstructural variability.
The embedded variability in initial hardening $\delta_{\kappa_0}$ and sub-threshold porosity $\delta_{\phi_0}$ appear to be uncorrelated.
The coefficient of variation of the initial hardening parameter $\kappa_0$ is approximately 0.25.
In the constitutive model this parameter is tied to characteristics of the initial dislocation network present prior to mechanical testing.
The coefficient of variation of the implicit porosity parameter $\phi_0$ is about 0.42.
The constitutive model interprets this parameter as indicative of the amount of preexisting, distributed, sub-threshold porosity and void-like defects.
These values indicate there is significant sample-to-sample variability in the process used to create the dogbones, at least as perceived through the chosen model.
It is also apparent that the predicted subthreshold porosity of 0.09 $\pm$ 0.04 is considerably larger than the visible porosity $\approx$ 0.008, which implies a large number of small diameter voids in this AM material.
These presumed voids seem to play a dominant role in the variability of plastic and failure characteristics of the AM material.


\section{Conclusion} \label{sec:conclusion}
In our efforts to characterize and model material variability with embedded uncertainty methods we have extended our previous work \cite{rizzi2019bayesian} to simultaneously handle sources of epistemic uncertainty due to limited experimental data and irreducible uncertainty due to the high dimension of microstructure configurations.
Specifically, we have developed a pseudo-marginal likelihood based method to handle and interpret the effects of explicit resolvable voids via mesh realizations and subthreshold, implicit porosity as described by the constitutive model and its parameters.
We were able to investigate its convergence properties with an inexpensive mechanics problem and showed that reasonable estimates can be made with relatively few realizations.
Furthermore, the proposed embedded uncertainty framework gives a distribution of response for an ensemble of samples that discerns not just uncertainty due to lack of data but also variability in physical parameters needed to explain variability in selected quantities of interest.
Thus it provides a means for physically characterizing the microstructural variability.
Using this method, we determined the influence of parameters and their contribution to variability of stresses and strains corresponding to yield, ultimate strength and failure of a large set of AM test specimens.
The methods we developed for characterizing controllable/parametric and uncontrollable/microstructural unknowns has direct relevance for design and process optimization.
For the present case of AM 17-4PH it appears the majority of the failure behavior is explained by the sub CT threshold ($< 7.5 \mu$m) porosity implicit in the selected constitutive model.

In developing the proposed framework for interpreting variability of microstructure through physical parameters a number of questions remain open.
While the selected constitutive model is appropriate for materials undergoing plasticity and damage, the consideration of alternate models by extension of the framework to model selection is one obvious next step.
Also a more rigorous Bayesian parameter selection process could be enabled with more computing power.
The validity of the explicit/implicit split treatment of voids could be explored further.
Experimental characterization at higher resolution or coarsening the current data and holding out the higher resolution information is another means of testing this hypothesis, albeit a challenging and expensive one. The generality of the proposed framework seems evident but extension to problems with other sources of microstructural variability are needed.

The conclusion that implicit porosity controls the observed variability, and the model components through which it is effected, is specific to the chosen additive manufacturing process, and quite possibly to the uniaxial tension tests. However, the embedded uncertainty methods developed here can be applied to materials and structures produced by any manufacturing process and undergoing deformation of any character. Against the backdrop of the framework developed here, further individual studies, while remaining limited in the generality of their conclusions, can add to a coherent body of UQ investigations of material variability. This sets up the possibility of encoding a more general body of engineering knowledge for design under uncertainty, enabling more precise quantification of margins and therefore more optimal designs. Current work by the authors seeks to uncover more universal relations in terms of correlations among observed material variability, inputs to constitutive models and specimen geometries. These studies will form the subject of a future communication.

\section*{Acknowledgments}
We are grateful for informative discussions with Kyle Karlson (Sandia).
This work was made possible by Sierra ({\tt sierradist.sandia.gov}), Dakota ({\tt dakota.sandia.gov}), and UQTk \linebreak ({\tt www.sandia.gov/UQToolkit/}).
This work was supported by the LDRD program at Sandia National Laboratories, and its support is gratefully acknowledged.
Sandia National Laboratories is a multimission laboratory managed and operated by National Technology and Engineering Solutions of Sandia, LLC., a wholly owned subsidiary of Honeywell International, Inc., for the U.S. Department of Energy's National Nuclear Security Administration under contract DE-NA-0003525. This paper describes objective technical results and analysis. Any subjective views or opinions that might be expressed in the paper do not necessarily represent the views of the U.S. Department of Energy or the United States Government.



\appendix \numberwithin{equation}{section}

\section{Karhunen-Lo\`eve Representation of the Porosity Process} \label{app:kle}

Given the experimental mean porosity $\meanporosity$ and spatial correlation of the observable voids, $\text{R}_{\porosity \porosity}(r)$, we construct a Karhunen-Lo\`eve (KL) representation of the porosity process follow the methodology proposed by Ilango \etal \cite{Ilango:2016}.
Specifically, we model the random porosity of the test specimens as a binary random process given by
\begin{equation}\label{disc_process}
\porosity(\xb) = \left \{ \begin{array}{ll}
1 \ , & \xb \in \text{void region} \\
0 \ , & \text{otherwise}
\end{array}\right .
\end{equation}
with mean porosity of $\meanporosity$ (\ie $\text{P}(\porosity=1) = \meanporosity$) and the variance of this process is $\meanporosity (1 - \meanporosity)$.
We assume that the process is homogeneous and possesses an isotropic two-point correlation function given by
\begin{equation}\label{disc_process_R2}
\text{R}_{\porosity\porosity} \left( \xb_1, \xb_2  \right) = \expectation \left[ \porosity \left( \xb_1 \right) \porosity \left( \xb_2 \right) \right] \ .
\end{equation}

Alternatively, we can model the normalized (zero-mean and unit-variance) process using the linear transform:
\begin{equation}
\Zc = \frac{\meanporosity - \porosity}{\sqrt{\meanporosity(1-\meanporosity)}} \ .
\end{equation}
The binary random process $\Zc(\xb)$ is constructed through an intermediate zero-mean, unit-variance Gaussian process $\Yc(\xb)$ modeled using KLE.
This involves transforming $\Yc$ (a continuous process) to $\Zc$ (a binary process) using the mapping:
\begin{equation}\label{y_to_z_trans}
\mathpzc{z} = \mathpzc{z}\left( y \right)= \left \{ \begin{array}{ll}
\frac{\meanporosity - 1}{\sqrt{\meanporosity(1-\meanporosity)}} \ , & y < y^* \\
\frac{\meanporosity}{\sqrt{\meanporosity(1-\meanporosity)}} \ , & {\rm otherwise}
\end{array}\right .
\end{equation}
with the threshold $y^*$ satisfying the identity \cite{Ilango:2016}:
\begin{equation}\label{y_to_z_trans2}
\frac{1}{\sqrt{2 \pi}} \int_{-\infty}^{y^*} e^{-t^2/2} \ dt = \meanporosity \ .
\end{equation}
chosen to preserve the mean porosity $\meanporosity$.

The correlation functions of processes $\Zc$ and $\Yc$ are related through the following identity
\begin{align}
\text{R}_{\Zc \Zc}( \xb_1, \xb_2) & = \expectation \left[ \Zc(\xb_1) \Zc(\xb_2) \right] \nonumber \\
& = \int \Zc(\xb_1) \Zc(\xb_2) \, {\rm p}_{_{\Zc \Zc}}( \Zc(\xb_1), \Zc(\xb_2)) \,  \mathrm{d}\Zc( \xb_1) \mathrm{d}\Zc(\xb_2)  \nonumber \\
& = \int \mathpzc{z}( \Yc( \xb_1) ) \mathpzc{z}( \Yc( \xb_2)) \, {\rm p}_{_{\Yc\Yc}}( \Yc(\xb_1), \Yc(\xb_2)) \,  \mathrm{d}\Yc(\xb_1) \mathrm{d}\Yc(\xb_2) \nonumber \\
& = \int \mathpzc{z}(y_1) \mathpzc{z}(y_2) \, {\rm p}_{_{\Yc\Yc}} (y_1, y_2) \,  \mathrm{d}y_1 \mathrm{d}y_2
\label{eq:Ryy_to_Rzz_trans_1}
\end{align}
with $y_1 = \Yc(\xb_1)$ and $y_1 = \Yc(\xb_2)$ used to simplify the exposition.
Since $\Yc$ is assumed to be a Gaussian random process with unknown correlation function $\text{R}_{\Yc\Yc}( \xb_1, \xb_2)$, we can write the joint \pdf, ${\rm p}_{_{\Yc\Yc}}( \Yc( \xb_1), \Yc( \xb_2))$, in terms of $\text{R}_{\Yc\Yc}$ as
\begin{align}
{\rm p}_{_{\Yc\Yc}}(\Yc(\xb_1), \Yc(\xb_2)) & = \frac{e^{-\left( \left( y_1^2 - 2 \text{R}_{\Yc\Yc} \left( \xb_1, \xb_2  \right)y_1y_2 + y_2^2\right) / \left( 2  \left( 1-\text{R}_{\Yc\Yc}^2 \left( \xb_1, \xb_2  \right) \right) \right) \right)}}{2 \pi \sqrt{1-\text{R}_{\Yc\Yc}^2 \left( \xb_1, \xb_2  \right)}}
\label{eq:pyy}
\end{align}

\eref{eq:Ryy_to_Rzz_trans_1} relates the known correlation function for the binary random process modeling the random porosity, $\text{R}_{\Zc \Zc} \left( \xb_1, \xb_2  \right)$, to that of the intermediate Gaussian random process $\text{R}_{\Yc\Yc} \left( \xb_1, \xb_2  \right)$.
Expanding the joint PDF ${\rm p}_{_{\Yc\Yc}} \left( \Yc(\xb_1), \Yc(\xb_2)  \right)$ in terms of Hermite polynomials as in \cite{Ilango:2016}
\begin{align}
{\rm p}_{_{\Yc\Yc}} \left( y_1, y_2 \right) & = \sum_{m = 0}^{\infty} \frac{\text{R}_{\Yc\Yc}^m}{m!} H_m \left( y_1 \right) H_m \left( y_2 \right) \varrho(y_1) \varrho(y_2)
\label{eq:pyy_exp}
\end{align}
with $H_m$ denoting the Hermite polynomial of order $m$ and $\varrho$ the standard Gaussian \pdf.
This expansion allows us to rewrite \eref{eq:Ryy_to_Rzz_trans_1} as \cite{Ilango:2016}
\begin{equation} \label{eq:Ryy_to_Rzz_trans_3}
\text{R}_{\Zc \Zc} \left( \xb_1, \xb_2  \right) = \sum_{m = 0}^{\infty} K_m^2 \text{R}_{\Yc\Yc}^m
\end{equation}
with
\begin{equation} \label{Km}
K_m = \frac{1}{\sqrt{m!}} \int \mathpzc{z} \left( y \right) H_m \left( y \right) \phi \left( y \right) dy \ .
\end{equation}
where we approximate  $K_m$ with Gauss-Hermite quadrature.
For this application, \eref{eq:Ryy_to_Rzz_trans_3} is truncated to 30 terms (chosen via a convergence study with results not included for brevity) and $\text{R}_{\Yc\Yc}$ is solved for numerically using a root-finding algorithm over different values of $\text{R}_{\Zc \Zc}$.

This procedure provides a mechanism to transform the two-point correlation of the normalized binary random process, $\Zc$, modeling the random porosity (obtained from observations) to that of the intermediate Gaussian random process, $\Yc$.
This intermediate Gaussian process was approximated by KLE as
\begin{equation}\label{Y_KLE}
\Yc( \xb ) \approx \displaystyle\sum_{j=1}^L \vartheta_j  \zeta_j ( \xb )
\end{equation}
where $\zeta_j=\sqrt{\lambda_j}f_j(x)$, and $\{\lambda_j, f_j\}_{j=1}^{L}$ are the first $L$-th eigenvalues and eigenfunctions of the correlation function $\text{R}_{\Yc\Yc}$ and the number of terms $L$ is optimally selected to capture  99.9\% of the energy (variance) of the process.
$\{{\vartheta_j}\}_{j=1}^{L}$ are independent standard Gaussian random variables.
For certain correlation functions, the eigenfunctions and eigenvalues are available in closed-form or semi-analytically (e.g. as numerical solutions to root-finding problems).
In general, however, one resorts to spatial discretization of the random process on the domain of interest that effectively maps it to a random vector (with the correlation function mapped to a covariance matrix) which can subsequently undergo a spectral decomposition to obtain the eigenvalue/eigenvector pairs.
See \cref{le2010spectral} for details.

\end{document}